\documentclass[conference]{IEEEtran}
\usepackage{fancyhdr}
\IEEEoverridecommandlockouts 

\usepackage[english]{babel}
\usepackage{newtxtext}
\usepackage{amsmath,amssymb,amsfonts}
\usepackage{newtxmath}
\usepackage{booktabs}
\usepackage{cite}
\usepackage{caption} 
\usepackage{subcaption}
\usepackage{enumitem}
\usepackage{graphicx}
\usepackage{xspace}
\usepackage{algorithm}
\usepackage{algpseudocode}
\usepackage[hidelinks]{hyperref}

\newcommand{\compactfigurespace}{\vspace{-0.6em}}
\newcommand{\compactstackedfigurespace}{\vspace{-1.2em}}

\algrenewcommand{\algorithmiccomment}[1]{\hfill\(\triangleright\) #1}

\newcommand{\subflows}{subflows\xspace}
\newcommand{\Subflow}{Subflow\xspace}

\newcommand{\rss}{RSS\xspace}
\newcommand{\lb}{load balancing\xspace}
\newcommand{\pfld}{PFLD\xspace}
\newcommand{\plb}{PLB\xspace}
\newcommand{\ecmp}{ECMP\xspace}
\newcommand{\encircle}[1]{%
  \scalebox{1.15}{\textcircled{\scriptsize #1}}%
}

\title{Ordered Chaos: Towards Adaptive Multipath Load Balancing with Precise Fast Loss Detection}

\author{
\IEEEauthorblockN{
Tommaso Bonato\IEEEauthorrefmark{1}\IEEEauthorrefmark{2}\textsuperscript{1},
Ahmad Ghalayini\IEEEauthorrefmark{2},
Nadeen Gebara\IEEEauthorrefmark{2},
Benoit Pit-Claudel\IEEEauthorrefmark{3}\textsuperscript{2},\\
Jiayong Li\IEEEauthorrefmark{1},
Mikhail Khalilov\IEEEauthorrefmark{1},
Luca Piccolboni\IEEEauthorrefmark{2},
Waleed Reda\IEEEauthorrefmark{2},\\
Pei-Huan Tsai\IEEEauthorrefmark{4}\IEEEauthorrefmark{2},
Tao Ji\IEEEauthorrefmark{2},
Abdul Kabbani\IEEEauthorrefmark{2},
Torsten Hoefler\IEEEauthorrefmark{1}\IEEEauthorrefmark{2}
}
\IEEEauthorblockA{\IEEEauthorrefmark{1}ETH Zurich, Zurich, Switzerland\\
\{tommaso.bonato,jiayong.li,mikhail.khalilov,torsten.hoefler\}@inf.ethz.ch}
\IEEEauthorblockA{\IEEEauthorrefmark{2}Microsoft Corporation, Redmond, United States of America\\
\{aghalayini,nadeengebara,lpiccolboni,waleedreda,taoji,abdulkabbani\}@microsoft.com}
\IEEEauthorblockA{\IEEEauthorrefmark{3}MIT, Cambridge, United States of America\\
bpitcla@alum.mit.edu}
\IEEEauthorblockA{\IEEEauthorrefmark{4}Columbia University, New York, United States of America\\
pt2630@columbia.edu}
}

\newcommand{\symmetricCaption}{Symmetric Topology}
\newcommand{\asymmetricCaption}{Asymmetric Topology}

\begin{document}

\maketitle

\thispagestyle{fancy}
\lhead{}
\rhead{}
\chead{}
\lfoot{\footnotesize{%
SC26, November 15-20, 2026, Chicago, Illinois, USA
\newline 979-8-3195-4789-7/26/\$31.00 \copyright 2026 IEEE}}
\rfoot{}
\cfoot{}
\renewcommand{\headrulewidth}{0pt}
\renewcommand{\footrulewidth}{0pt}

\begingroup
\renewcommand{\thefootnote}{\arabic{footnote}}
\footnotetext[1]{Tommaso Bonato has since left Microsoft.}
\footnotetext[2]{Benoit Pit-Claudel has since joined Amazon. This work does not relate to his position at Amazon.}
\endgroup
\setcounter{footnote}{2}

\begin{abstract}\label{sec:abstract}Load balancing in best-effort datacenter fabrics faces a tradeoff. Order-preserving schemes such as Equal-Cost Multipath (ECMP) retain loss semantics but use path diversity coarsely, while packet spraying balances load well but removes the ordering signals needed for fast, precise loss detection. We present Ordered Chaos, an end-host design that preserves enough structure for reliable loss detection without giving up fine-grained multipath routing. Reactive Subflow Spraying (RSS) divides each flow into ordered, dynamically routed subflows. Precise Fast Loss Detection (PFLD) combines this ordering with probes to distinguish reordering from congestion, corruption, tail, and retransmission losses within one RTT. Across simulations, AI and HPC workloads, and a DPU prototype, RSS+PFLD improves tail flow completion time by up to 3$\times$, eliminates up to 99\% of timeout-based recoveries, and accelerates collective and training communication by up to 35\%.
\end{abstract}

\section{Introduction}
Large-scale AI training and tightly-coupled HPC workloads occupy tens of thousands of accelerators for weeks to months, making the transport layer's ability to detect and recover from packet losses a direct factor in time-to-result and cluster efficiency. At scale, these environments experience packet loss due to transient congestion, tail drops, failures, and persistent packet corruption arising from aging optics, hardware faults, and link degradation.
\begin{figure}[!t]
  \centering
  \includegraphics[width=\linewidth]{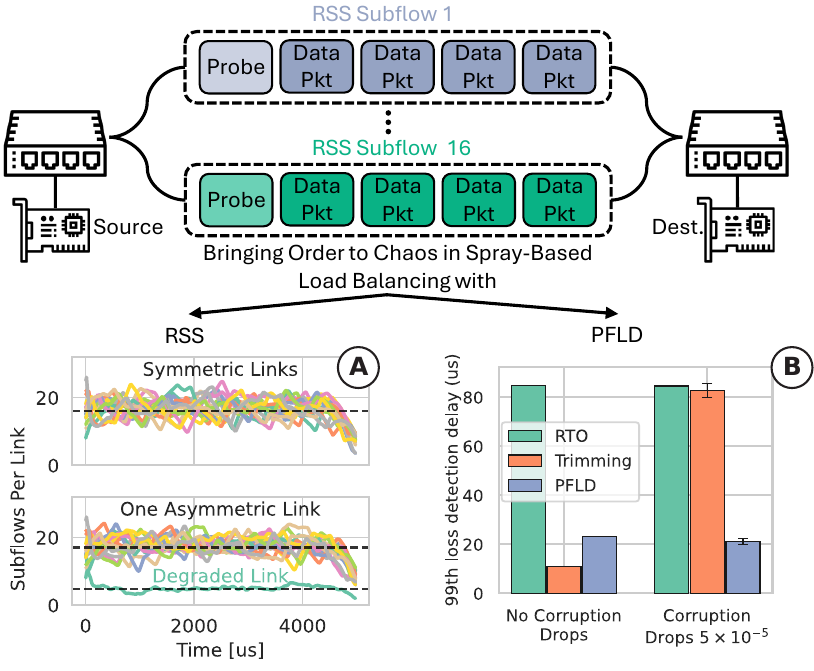}
  \caption{Scheme showcasing RSS+PFLD with 16 subflows. \protect\encircle{A} shows the performance of RSS during a tornado pattern with a symmetric and an asymmetric topology. \protect\encircle{B} shows the time it takes to detect a packet loss for different schemes.}
  \label{fig:top_fig}
  \vspace{-1em}
\end{figure}
Production measurements report that inter-switch and inter-card loss events, which include silent packet drops and packets discarded after link corruption, account for about 18\% of packet drops associated with network performance anomalies~\cite{zhou2020flowevent}, and end-to-end loss ratios in the range of $10^{-5}$ to $10^{-4}$ have been consistently reported within datacenters~\cite{guo2015pingmesh, zhuo2017corruption, joshi2023linklocal, roy2017passive}. These losses compound across multi-hop paths, stalling collectives, desynchronizing bulk-synchronous phases, and increasing training costs~\cite{alibaba}. Modern datacenter best-effort (lossy) fabrics expose multiple parallel paths between nodes~\cite{hammingmesh, slingshot, facebook_sigcomm} to meet throughput and resiliency demands, yet turning this path diversity into consistent application performance in the presence of loss and reordering, without introducing new reliability problems, remains an open problem~\cite{plb, prestoLB, bonato2025repsrecycledentropypacket, mptcp, flowlet, flowbender, drill, sglb, spray, conga, conweave, bonato2025flowcutswitchinghighperformanceadaptive}.

In practice, most datacenter fabrics attempt to exploit this path diversity using per‑flow load‑balancing mechanisms, with Equal‑Cost Multipath (\ecmp{}) serving as the dominant starting point. \ecmp{} hashes a flow’s five‑tuple to select a single path for its lifetime. While simple and order‑preserving, \ecmp{}’s hash‑based static mapping can lead to collisions that oversubscribe some links while under‑utilizing others, and its lack of adaptivity prevents effective response to asymmetric conditions, congestion, or failures~\cite{hedera, prestoLB, hashing_polarization}. Reactive flow‑level schemes such as \plb{}~\cite{plb} reduce persistent collisions through occasional rerouting, but still operate at coarse, per‑flow granularity, limiting their ability to fully utilize available path diversity.

To overcome these limitations, the ecosystem has increasingly converged on multipath spraying, shifting from per‑flow path selection to much finer‑grained packet‑level traffic distribution across paths. Correspondingly, industry efforts such as Ultra Ethernet~\cite{ultraethernet2025spec, hoefler2025ultraethernetsdesignprinciples}, Falcon~\cite{falcon}, SRD~\cite{srd}, and the NVIDIA ConnectX-8~\cite{nvidia_connectx8_supernic_um_2025} and BlueField-3~\cite{nvidia_bluefield3_dpu_2021_datasheet} NICs all adopt out-of-order-capable transports with per-packet spraying or other fine-grained \lb{} mechanisms.

However, packet spraying destroys the ordering signals that traditional transports rely on for fast loss recovery: an endpoint cannot immediately distinguish between a true loss and a delayed packet. To manage this without the prohibitive memory cost of per-packet timers, modern DPUs like BlueField-3 employ hardware-resident bitmaps and per-Queue Pair (QP) gap timers~\cite{nvidia_bluefield3_dpu_2021_datasheet}. While these end-to-end timers eventually detect all losses, including those caused by corruption, they do so without the benefit of explicit network signaling. Consequently, some recent systems are starting to rely on switch-side packet trimming for near-instant congestion notification~\cite{nvidia_spectrumx_platform_2025, htsim}. Critically, because corruption drops typically occur at the PHY/MAC layer before a packet is queued, the switch cannot generate a trim notification. This leaves best-effort transports to rely on the expiration of conservative hardware timeouts, which, while functional for detection, introduce significant tail latency ($P99$) spikes compared to the rapid recovery seen in congestion-induced drops.

Based on these observations, we propose \textbf{Ordered Chaos}, a unified, host-based \lb{} and loss-recovery system for AI training and tightly-coupled HPC collectives. The key insight behind Ordered Chaos is that preserving \emph{just enough ordering structure} enables fast, deterministic loss signaling without sacrificing the utilization benefits of fine-grained multipathing. Building on this insight, Ordered Chaos introduces two \emph{novel and synergistic contributions}: \emph{Reactive Subflow Spraying (\rss{})} and \emph{Precise Fast Loss Detection (\pfld{})}.

\rss{} introduces a new operating point between per-flow hashing (e.g., \ecmp{}/\plb{}) and per-packet spraying (e.g., RPS) by deterministically splitting each flow into $\sigma$ logical subflows via round-robin mapping, such that each subflow carries a predictable packet subsequence. Rather than reacting at flow granularity, \rss{} periodically reroutes only the worst-performing subflow based on end-host congestion signals (e.g., RTT or ECN), achieving fine-grained adaptivity under asymmetry and failures while intentionally preserving in-order delivery within each subflow. \pfld{} exploits this preserved structure to enable a novel, \emph{receiver-driven} loss recovery mechanism: by combining per-subflow sequence gaps with lightweight probes, \pfld{} deterministically distinguishes reordering from true loss and enables prompt recovery from congestion, corruption, tail, and retransmission losses without switch support or unbounded per-packet state. This receiver-driven, deterministic loss identification enables precise recovery actions and substantially reduces spurious or duplicate retransmissions compared to time-based, heuristic approaches.

Figure~\ref{fig:top_fig} illustrates how \rss{} preserves partial in‑order delivery across subflows and how PFLD leverages this structure for fast and precise loss recovery.
\protect\encircle{A}~shows RSS on a symmetric and an asymmetric fat-tree topology with 16 subflows. In the asymmetric case, one of the 16 output links is throttled to 20\% capacity; RSS reacts within microseconds, redistributing subflows so the degraded link carries proportionally fewer, converging to the ideal distribution (dotted lines). \protect\encircle{B}~compares the 99th-percentile loss detection delay during an incast workload. With corruption drops ($\approx 10^{-5}$), RTO‑based recovery takes $60$–$80\,\mu s$, and packet trimming cannot detect corruption drops because no trimmed header is produced. PFLD detects both congestion and corruption losses within a single RTT ($<20\,\mu s$). We define \emph{corruption drops} as losses not caused by congestion.

\begin{figure*}[!t]
    \centering
    \includegraphics[width=0.95\textwidth]{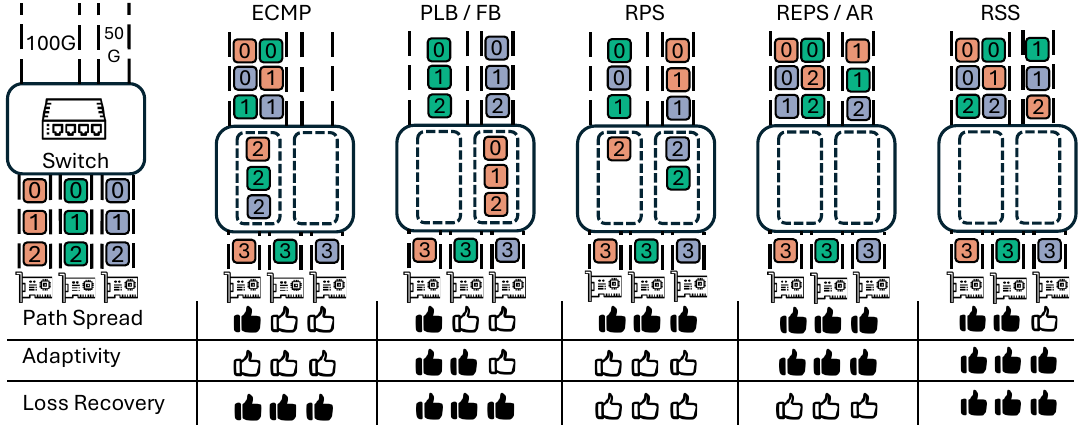}
    \caption{Routing decisions for different \lb approaches using a sample switch with 3 50 Gbps input ports and 2 output ports (one running at 100 Gbps and the other at 50 Gbps). Packets flow upward, from the input ports at the bottom to the output ports at the top. \textit{FB} stands for \textit{FlowBender} and \textit{AR} for \textit{Adaptive Routing}.}
    \label{fig:comparison}
    \vspace{-1em}
\end{figure*}

Overall, without any switch modifications, Ordered Chaos reduces the non-overlapped communication time of LLaMA-7B training by 28\%, speeds up ring all-reduce by 35\%, and improves tail flow completion time by up to 3$\times$ across AI collectives, HPC applications (ICON, LAMMPS, MILC), and microbenchmarks. Our main contributions are:

\begin{itemize}
 \setlength\itemsep{0mm}
 \item \textbf{Reactive Subflow Spraying (\rss{}).} We introduce \rss{}, a subflow-level reactive load-balancing mechanism that partitions each flow into a small number of strictly ordered subflows and periodically reroutes only the worst-performing subflow based on simple end-host signals (e.g., RTT or ECN). \rss{} achieves near-uniform link utilization under asymmetric and dynamic conditions while requiring no switch modifications. 

\item \textbf{Precise Fast Loss Detection (PFLD).} We present \pfld{}, a receiver-driven loss detection mechanism that leverages \emph{subflow-level ordering} to precisely detect and recover from congestion, tail, retransmission, and corruption losses without packet trimming. PFLD combines lightweight probe packets with packet-sequence-number (PSN)-aware entropy encoding to deliver fast and deterministic loss signals with minimal memory requirements. 

\item \textbf{Unified transport co-design with minimal state.} We co-design RSS and PFLD so that ordered subflows provide the structural signal needed for precise end-host loss detection, while accurate and timely loss feedback enables RSS to react aggressively without hardware assistance. This tight integration keeps per-flow state small (a few per-subflow counters and a receiver bitmap), ensuring scalability and robustness in the presence of asymmetry, failures, and corruption drops.

 \item \textbf{Evaluation at scale and on hardware.} 
We evaluate \rss{}+\pfld{} at scale using large‑scale packet‑level simulations over a best‑effort underlying fabric, spanning AI collectives, end‑to‑end LLM training traces, and replayed HPC workloads, and further validate implementation feasibility via a programmable DPU prototype. Ordered Chaos improves tail flow completion time by up to $3\times$ and accelerates collective and training communication by up to 35\%.
\end{itemize}

\section{Motivation} \label{motivation}

Multipath, out-of-order-capable transports are becoming the prevailing design choice, yet a fundamental tension remains unresolved. Existing multipath load balancing and transport schemes struggle to simultaneously deliver fine‑grained load balancing, fast loss detection for both congestion‑ and corruption‑induced drops, and end-host only deployability. Figure~\ref{fig:comparison} summarizes the dominant multipath load‑balancing approaches used in today’s transports and highlights this tension: order‑preserving designs such as ECMP and FlowBender (FB) underutilize path diversity and react slowly to asymmetry, whereas packet‑spraying schemes such as Random Packet Spraying (RPS) and its adaptive variants (REPS/AR) achieve high utilization by sacrificing the ordering signals required for timely loss detection.

We further observe that, perhaps counter‑intuitively, per‑packet spraying is not necessarily needed to achieve high load‑balancing effectiveness (Section~\ref{micro_eval}). \rss{} leverages this insight by operating at subflow granularity, retaining most of the utilization benefits of packet‑level spreading while preserving controlled ordering. At the same time, although packet trimming can rapidly signal congestion‑induced losses, it provides no notification for corruption drops, which occur before queueing. This limitation motivates the combination of ordered subflows (\rss{}) with fast, end‑host loss detection (\pfld{}).
In the remainder of this section, we characterize the prevalence and impact of corruption drops in production and explain why packet trimming, despite its effectiveness for congestion, is insufficient for fast loss recovery in this setting.

\subsection{Corruption Drops in Production} \label{failures}

Large-scale AI training and HPC jobs run for days to months across tens of thousands of links. At this scale, both systemic failures (link flaps, switch failures, routing changes) and random corruption drops are the norm: production AI clusters report 5K to 60K link-flap incidents per day~\cite{alibaba}, and corruption drops caused by aging optics, bit errors, connector degradation, and ASIC glitches occur continuously across such large deployments~\cite{facebook_sigcomm}.

Production measurements report that inter-switch and inter-card loss events, which include silent packet drops and packets discarded after link corruption, account for about 18\% of packet drops associated with network performance anomalies. These events take about 161 minutes to locate on average and account for half of the packet drops whose localization takes more than 180 minutes~\cite{zhou2020flowevent}.

Corruption compounds across multi-hop paths. If the per-link corruption probability is $p_{\text{link}}$ and a path traverses $h$ links, the end-to-end failure rate is
\[
p_{\text{end}} = 1 - (1 - p_{\text{link}})^{h} \approx h \cdot p_{\text{link}} \quad \text{for small } p_{\text{link}}.
\]
For instance, with a per-link corruption rate of $5 \times 10^{-5}$ (the rate we adopt for our evaluation, consistent with reported values~\cite{zhuo2017corruption, uno}) and a typical 6-hop datacenter path, the end-to-end loss probability is roughly $3 \times 10^{-4}$. This aligns with measurement studies that consistently report end-to-end loss ratios in the $10^{-5}$ to $10^{-4}$ range under normal operation~\cite{zhuo2017corruption, guo2015pingmesh, roy2017passive, joshi2023linklocal}, with localized degradations transiently pushing affected links into the $10^{-3}$ to $10^{-2}$ range before quarantine or repair.

These failure rates mean the network is almost always experiencing some fault or recovery event. Transports must remain effective under persistent change rather than only in steady state, making robust loss detection and recovery mechanisms critical for maintaining cluster efficiency~\cite{alibaba}.

\subsection{Why Trimming Is Not Enough} \label{sec:why_trimming}

To mitigate packet loss in the presence of reordering, packet trimming has been proposed~\cite{htsim, cp}: when a switch queue is full or reaches a threshold, the incoming data packet's payload is stripped and the remaining header is pushed into a high-priority queue. The receiver sends a NACK upon receiving a trimmed header, prompting the sender to retransmit. Although trimming is being adopted in several recent best-effort protocols~\cite{srd, ultraethernet2025spec}, due to its effectiveness in rapidly signaling congestion‑induced losses, it has two fundamental limitations. First, many production deployments still lack trimming support entirely~\cite{lossyrdma}. Second, and more critically, corruption drops occur \emph{before} a packet is queued at the switch, leaving no trimmed header to signal the loss. In such cases, transports without end‑host loss detection must fall back to retransmission timeouts, delaying recovery by orders of magnitude relative to the round‑trip time and directly increasing job completion times.

Credit-based flow control similarly prevents congestion drops but not corruption drops that occur at the PHY/MAC layer before queueing. On such lossless fabrics, PFLD's role narrows to detecting corruption and link failures, while RSS remains useful when the fabric exposes endpoint-controlled path selection.

\section{RSS Design}\label{sec:design}

The \rss{} sender logic is detailed in Algorithm~\ref{alg:main} and its defining characteristics are described next.

\begin{figure}[bt]
  \centering
  \includegraphics[width=\linewidth]{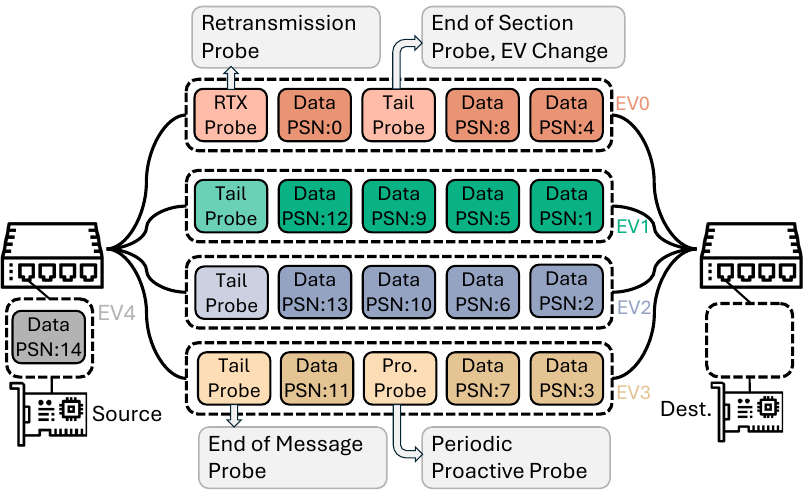}
  \caption{Detailed scheme of RSS+PFLD with 4 subflows. In the example we assume \textit{PSN 0} is lost before the visualization. We also assume that each \textit{EV} leads to a different path.}
  \label{fig:pfld_overall}
  \vspace{-1em}
\end{figure}

\subsection{\Subflow Spraying}\label{subsec:spraying}

The \rss{} mechanism deterministically partitions a single flow into $\sigma$ parallel subflows each representing a $1/\sigma$ fraction of the aggregate bandwidth. This can be seen in Figure~\ref{fig:pfld_overall} which jointly illustrates the operation of \rss{} and \pfld{} with $\sigma=4$. The parameter $\sigma$ serves as a programmable tuning knob for load-balancing granularity: while $\sigma=1$ reverts to static ECMP, higher values approach the high-entropy benefits of per-packet spraying. In this work, we fix $\sigma=16$ to balance path diversity with reordering overhead in high-radix fabrics.

Unlike MPTCP~\cite{mptcp}, which necessitates independent state machines per subflow, \rss{} utilizes a shared flow state across all paths. Furthermore, while MPTCP achieves load balancing through fine-grained rate modulation ($bits/s$), \rss{} employs a \textit{discrete steering} approach. By dynamically adjusting the number of subflows allocated to a specific physical path (see \ref{subsec:reactive}), \rss{} provides reactive load balancing without the overhead of hardware rate-shapers or per-subflow timers.

To ensure compatibility with existing network fabric, \rss{} steers subflows by manipulating the entropy value (EV) in packet headers to leverage standard ECMP hashing at switches. Consistent with prior work~\cite{bonato2025repsrecycledentropypacket,plb}, we utilize the 16-bit source port, the IPv6 flow label~\cite{plb}, or the Ultra Ethernet entropy field~\cite{ultraethernet2025spec} as the entropy source. 

While subflow and flowlet abstractions are established in protocols like MPTCP~\cite{mptcp} and adaptive routing schemes~\cite{conga}, \rss{} introduces a \textit{deterministic mapping paradigm} that eliminates the ambiguity inherent in traditional grouping. Unlike FlowBender's~\cite{flowbender} reactive, flow-level switching or the stochastic nature of gap-dependent flowlets, \rss{} provides a predictable, hardware-native sequence-to-path correspondence. By proactively partitioning the flow regardless of application-level ``gaps,'' \rss{} creates the architectural foundation for PFLD, enabling fine-grained, low-latency loss detection without the overhead of tracking probabilistic reordering events.

\begin{algorithm}[t]
\footnotesize
\captionsetup{font=small}
\caption{\small Pseudocode for an \rss{} sender using mean RTT.}
\label{alg:main}
\begin{algorithmic}[1]
\Require $\sigma$: number of RSS subflows; $N=\log_2(\sigma)$

\Procedure{updateEv}{sid}
    \If{PSN-based PFLD is enabled}
        \State ev[sid] $\leftarrow$ low16(nextPsn(sid))
    \Else
        \State ev[sid] $\leftarrow$ (rand() $\ll$ N) + sid
    \EndIf
\EndProcedure

\Procedure{onAck}{ack}
    \State sid $\leftarrow$ getSubflowId(ack)
    \State rttSum[sid] $\leftarrow$ rttSum[sid] + ack.rtt
    \State samples[sid] $\leftarrow$ samples[sid] + 1 \Comment{RTT samples this interval}

    \If{now $\ge$ nextRerouteTime}
        \State nextRerouteTime $\leftarrow$ now $+\tau+$ randomJitter()
        \State active $\leftarrow \{i \mid$ samples[$i$] $> 0\}$
        \If{active $\ne \emptyset$}
            \State sid $\leftarrow$ worstMeanRtt(rttSum, samples, active)
            \State \textsc{updateEv}(sid)
        \EndIf
        \State reset(rttSum, samples)
    \EndIf
\EndProcedure

\Statex

\Procedure{onPacketSchedule}{seqNo}
    \State \Return ev[seqNo $\bmod$ $\sigma$] \Comment{Round-robin subflow selection}
\EndProcedure

\end{algorithmic}
\vspace{-0.4em}
\end{algorithm}

In Algorithm~\ref{alg:main}, \textsc{updateEv} uses two EV encodings. For the power-of-two $\sigma$ used here, let $N=\log_2(\sigma)$. Plain RSS randomizes the upper bits to select a new path while retaining the subflow ID in the low $N$ bits. PSN-based PFLD instead uses the low 16 bits of the first PSN after a reroute; because subflow $s$ carries PSNs $p$ for which $p \bmod \sigma=s$, the low $N$ bits still identify the subflow while the EV also marks the start of its new ordered section (Section~\ref{pfld}).

\subsection{Reactive Rerouting}\label{subsec:reactive}
While \plb~\cite{plb} is limited to serial probing of a single path, \rss{} concurrently monitors $\sigma$ paths to maintain a comprehensive view of network state. Unlike gap-dependent flowlet switching~\cite{flowlet}, which fails under the dense traffic typical of RDMA and AI collectives~\cite{mprdma}, \rss{} evaluates rerouting on a fixed schedule $\tau$ independent of traffic patterns. At the first feedback event at or after each deadline, the sender evaluates path-level metrics, such as average RTT or ECN marking fractions, and migrates the worst-performing subflow for which it collected samples. Since the EV determines the switch-level routing, updating this field enables near-instantaneous rerouting to a new physical path. This parallel, scheduled approach ensures high-radix fabric utilization even during periods of maximum link saturation. 

The update interval $\tau$ governs the stability and convergence of \rss{} as a control system. At flow creation, \textit{nextRerouteTime} is initialized to the first update deadline. Algorithm~\ref{alg:main} advances subsequent deadlines by $\tau$ plus a small random jitter. Path feedback is collected throughout each interval and reset after the rerouting decision. We recommend configuring $\tau$ as a function of the network RTT to ensure sufficient feedback, such as ECN marks or sample RTTs, is received before rerouting decisions are made. The jitter prevents global synchronization between independent flows, which can lead to periodic fabric oscillations. Setting $\tau$ too low risks rerouting from too few samples, while an excessively large $\tau$ increases convergence time and delays the response to transient hotspots. A deployment that must react to complete feedback loss can additionally schedule a no-feedback timer for each subflow.
We demonstrate the impact of the choice of $\tau$ in Section~\ref{sensitivity}.

\subsection{\rss{}'s Deployability}\label{subsec:requirements}
\rss{} requires only end-host support; no changes to network devices are needed. Recent work likewise demonstrates endpoint-controlled, path-aware load balancing with standard Ethernet mechanisms in low-diameter networks~\cite{bonato2026spritz}. It requires an end-host-settable path selector and endpoint support for out-of-order delivery across selectors while preserving order within each selector. Ethernet/RoCE exposes path selection through fields such as the UDP source port or IPv6 flow label, while Ultra Ethernet provides an entropy field~\cite{plb,bonato2025repsrecycledentropypacket,hoefler2025ultraethernetsdesignprinciples}. Slingshot, InfiniBand, and Omni-Path can provide the same abstraction when the NIC exposes equivalent path handles or entropy controls. The design is not applicable when a fabric hides all path choice or enforces strictly in-order endpoint delivery. Out-of-order-capable stacks are now available from cloud providers~\cite{srd,ultraethernet2025spec,nvidia_spectrumx_platform_2025} and in research prototypes~\cite{juggler}.

\rss{} has a small, $O(\sigma)$ memory footprint. The implementation maintains an entropy value, feedback count, accumulated RTT, maximum RTT, and ECN count per subflow, plus the next-update timestamp. A hardware implementation can instantiate only the accumulators required by its selected metric; the mean-RTT configuration in Algorithm~\ref{alg:main}, for example, needs only the entropy, feedback count, and accumulated RTT vectors.

\section{\pfld{} Design} \label{pfld} \label{sec:pfld}

While \rss{} results in out-of-order delivery \textit{across} subflows, it preserves \textit{in-order delivery} within each subflow. Since packets within a subflow sharing an EV are guaranteed to traverse the same physical path (Section~\ref{subsec:reactive}), they maintain strict sequential ordering relative to their specific subflow. \pfld{} leverages this deterministic structure for high-fidelity loss detection.

\subsection{Design Space and PSN-Based PFLD}

We considered several PFLD variants with progressively smaller state requirements, summarized in Table~\ref{tab:pfld_state}. An \emph{Infinite Memory} design tracks every in-flight PSN per EV, detecting gaps immediately but requiring state proportional to the bandwidth-delay product, which is prohibitive on NICs. A \emph{Fixed Path} variant constrains each subflow to a static EV with an arithmetic PSN subsequence, reducing state to one expected-PSN counter per path plus a single 512-bit per-flow bitmap; however, it cannot support rerouting. A \emph{Generation-Based} extension adds expected-PSN counters for each concurrent generation of every subflow while retaining one per-flow bitmap.

\textbf{PSN-Based PFLD} is our proposed state-optimal scheme that eliminates the need for explicit generation tracking. The key insight is to encode the start of each rerouted section directly in the EV. When RSS reroutes a subflow, it sets the EV to the low 16 bits of the next PSN scheduled on that subflow. The first packet on the new path carries this EV. Because the value is derived from that packet's PSN rather than chosen randomly, the receiver can recover the section's start PSN without a separate handshake or explicit generation state. A single 512-bit bitmap per flow records the received status of each packet relative to this start. Because round-robin spraying maps PSN $p$ to subflow $p \bmod \sigma$, the lower $\log_2(\sigma)$ EV bits still identify the subflow. With $\sigma=16$, the remaining 12 bits provide 4096 EVs per subflow. Generation PSNs vary across flows, message sizes, and reroutes; if two EVs collide on a poor path, RSS can reroute the affected subflow again.

\begin{figure}[t!]
  \centering
  \includegraphics[width=0.80\linewidth]{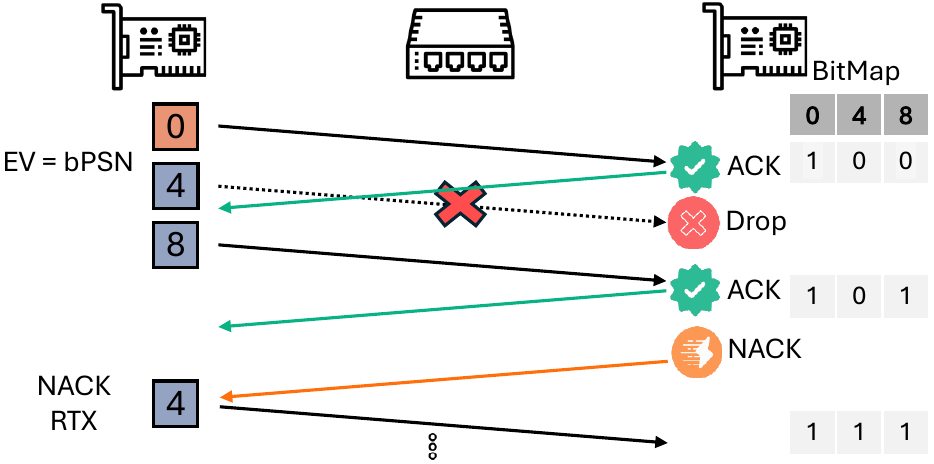}
  \caption{Operation principle of PSN-Based PFLD with 4 subflows. Only the first subflow is shown.}
  \label{fig:pfld3}
  \compactfigurespace
\end{figure}

\begin{table}[h]
\centering
{\footnotesize
\begin{tabular}{cc}
\toprule
\textbf{Version} & \textbf{State to Store} \\ \midrule
Infinite Memory   & \#BDP Packets $\times$ 32b PSN \\
Fixed Path        & $\#Paths \times 32\text{b ePSN} + 512\text{b bitmap}$ \\ \midrule
Gen-Based         & $\#Paths \times \#Gen \times 32\text{b ePSN} + 512\text{b bitmap}$ \\
PSN-Based         & 512b bitmap \\ \bottomrule
\end{tabular}
}
\caption{PFLD variants and their memory footprint. Bitmaps enable selective NACK; dropping them implies go-Back-N.}
\label{tab:pfld_state}
\end{table}

In the receiver-driven variant (Figure~\ref{fig:pfld3}), PSNs are global: subflow $s$ carries exactly those PSNs for which $p \bmod \sigma=s$, and each EV identifies the start of one ordered section of that subsequence. The receiver lifts the 16-bit EV into its current absolute-PSN window, computes the packet's bitmap offset, and marks the corresponding bit. Packets between the maximum received PSN index and the first unset bit are identified as lost and NACKed immediately for retransmission. Upon receiving NACKs, the sender enqueues the missing packets for retransmission.

Data placement is identical to RPS or another out-of-order transport and requires no separate reorder buffer: each packet carries its PSN and offset, so an out-of-order-capable RDMA stack places the payload directly at its final location. PFLD's per-flow bitmap is used only for loss detection. Running PFLD at the receiver provides two benefits. First, the receiver is the first endpoint with positive evidence of loss: a later packet or probe has arrived on the same ordered subflow while an earlier PSN has not. Second, it can immediately NACK the precise gap. A sender sees only missing ACKs, which may instead reflect ACK coalescing, ACK loss, or reverse-path delay and therefore require additional feedback or time-based thresholds such as RACK-TLP.

The required bitmap size is determined by the maximum unresolved PSN span in an active section and scales with rate, RTT, and packet size. At $400$\,Gb/s and $10\,\mu$s RTT, the NIC-wide BDP spans about $122$ packets with a $4$\,KiB maximum transmission unit (MTU) and $489$ packets with 1\,KiB packets. We therefore use a 512-bit bitmap, which covers both cases; this choice sizes the tracking window rather than constraining PFLD correctness to a particular MTU. For 512-byte packets or larger BDPs, a deployment can use a larger bitmap, such as 1024 bits, or overlapping generations without changing the detection logic. When RSS reroutes a subflow, the sender emits a tail probe on the old EV to close the section as shown in Figure~\ref{fig:pfld_overall}. The bitmap is indexed by absolute PSN and slides forward monotonically, so late-arriving packets on the old path are marked within the existing window. NACKs are scoped to each section's PSN range, preventing cross-section spurious retransmissions.

\begin{figure}[htbp]
    \centering
    \includegraphics[width=0.95\linewidth]{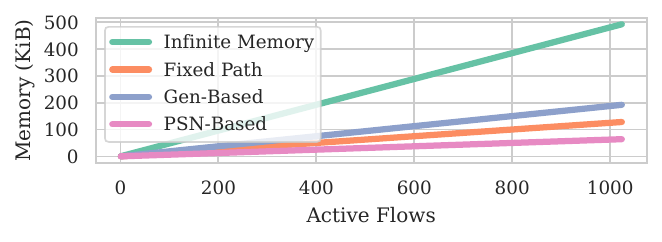}
    \caption{PFLD Memory Consumption Per-NIC.}
    \label{fig:memory}
\end{figure}

Figure~\ref{fig:memory} shows the memory occupancy per NIC for all variants assuming a 400\,Gb/s network with a 10\,$\mu$s RTT and a 4\,KiB MTU. PSN-Based PFLD requires less than 100\,KiB even at 1000 active flows, while the Infinite Memory variant reaches about 480\,KiB. The bitmap size can be adjusted based on deployment requirements and is not tied to the core RSS or PFLD design.

\subsection{Probing for Full Reliability}

PSN-based loss detection identifies drops whenever a subsequent packet arrives on the same subflow and reveals a gap. However, three corner cases require additional mechanisms: tail losses (no subsequent packet exists), prolonged loss bursts (detection waits indefinitely), and retransmission losses (non-sequential PSNs break gap detection). We address all three with lightweight probe packets shown in Figure~\ref{fig:pfld_overall}.

\begin{figure}[htbp]
    \centering

    \begin{subfigure}{0.48\linewidth}
        \centering
        \includegraphics[width=\linewidth]{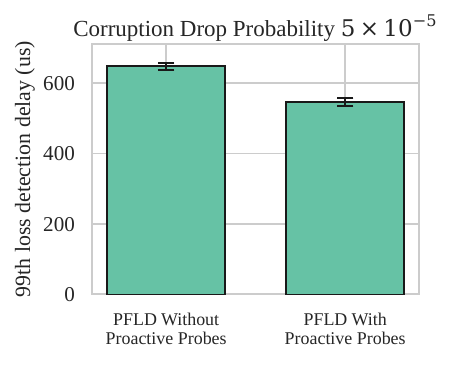}
        \caption{PFLD with proactive probes on and off during an incast.}
        \label{fig:probe_why}
    \end{subfigure}\hfill
    \begin{subfigure}{0.48\linewidth}
        \centering
        \includegraphics[width=\linewidth]{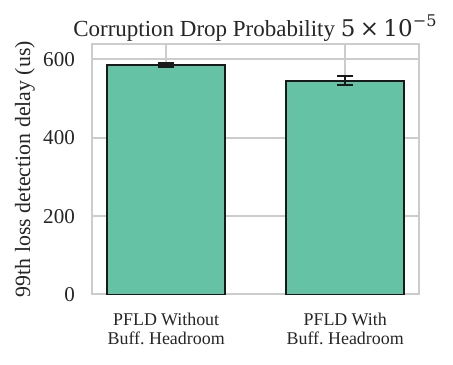}
        \caption{PFLD with and without dedicated buffer headroom.}
        \label{fig:headroom_why}
    \end{subfigure}

    \caption{Impact of probes on PFLD at corruption probability $5\times10^{-5}$.}
    \label{fig:probe_impact}
    \compactfigurespace
\end{figure}

\begin{figure*}[t]
     \centering
     \begin{subfigure}[b]{0.49\textwidth}
        \centering
        \includegraphics[width=\linewidth]{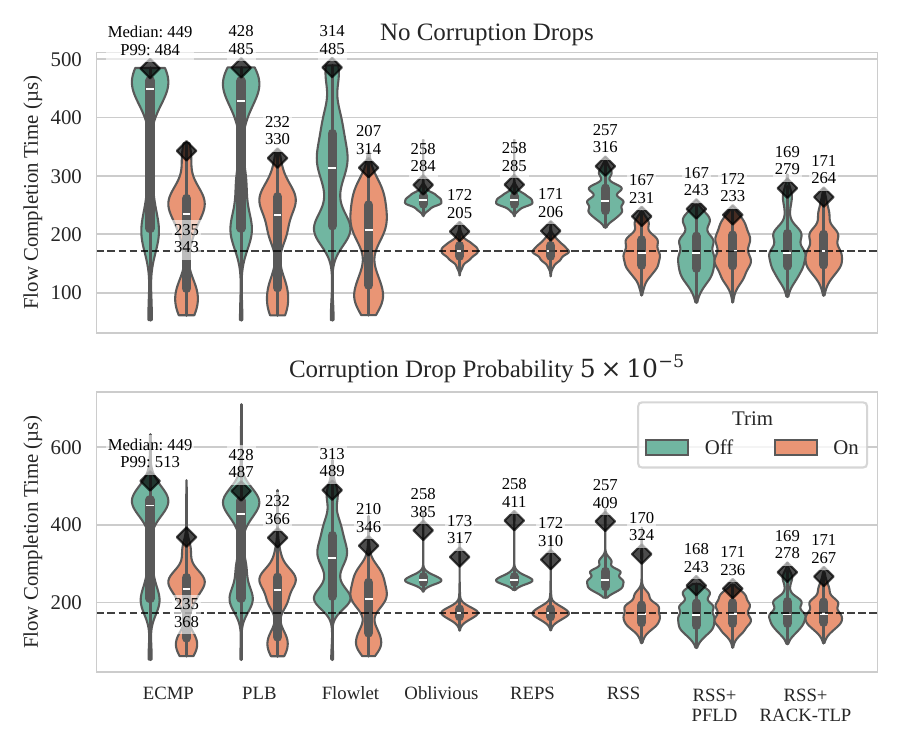}
        \caption{1 MiB 8:1 Incast}
        \label{fig:incast}
     \end{subfigure}
     \hfill
     \begin{subfigure}[b]{0.49\textwidth}
        \centering
        \includegraphics[width=\linewidth]{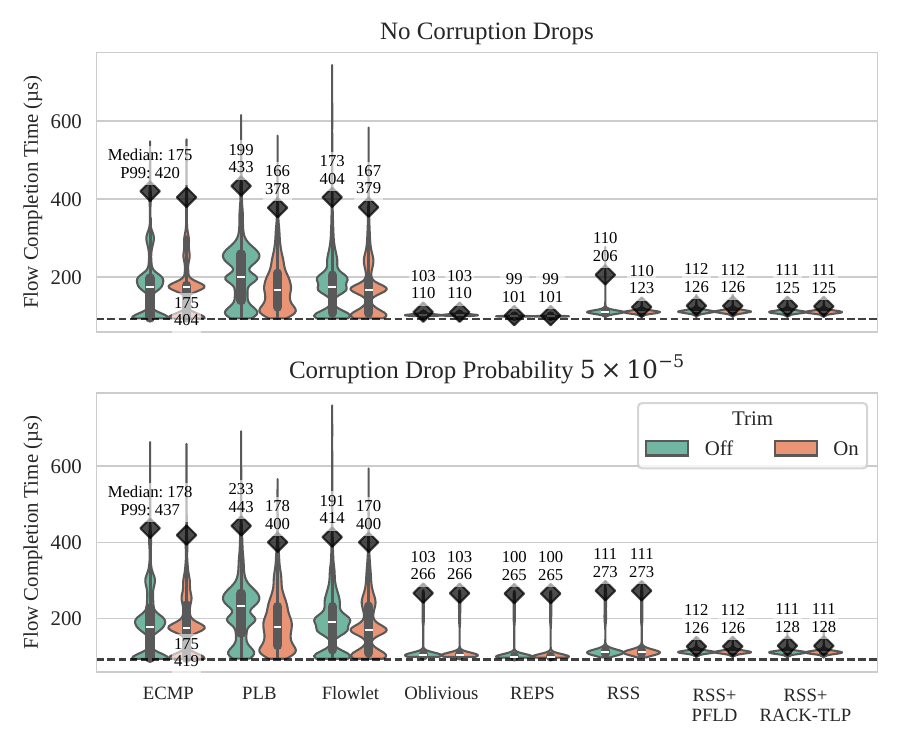}
        \caption{4 MiB Tornado Pattern}
        \label{fig:tornado}
     \end{subfigure}
    \caption{Microbenchmarks comparing different schemes with and without corruption drops. Positive-drop panels use corruption probability $5\times10^{-5}$. For each scheme, we also evaluate with and without packet trimming. The dashed line indicates the ideal FCT and the rhombus marker the 99th percentile.}
    \label{fig:micro}
\end{figure*}

\subsubsection{Probe Packet for Tail Loss}
The PSN-based gap mechanism alone cannot detect the loss of the last packet before an EV change or the last packet of a subflow, because no subsequent packet exists to reveal the gap. PFLD handles these cases with tail probes. With RSS, each message has multiple tail packets (one per subflow, plus additional ones at each reroute), so the sender appends a header-only \textit{tail probe} after the last data packet on each subflow. The probe carries the last PSN that should have been received before it, allowing the receiver to NACK any missing PSNs up to and including that value.

\subsubsection{Proactive Probing}
The gap-based mechanism only detects losses after receiving a packet. If a burst of consecutive packets is lost, detection is delayed until after the RTO, causing spurious retransmissions. The sender prevents this with periodic \textit{proactive probes}: header-only, low-priority packets that ensure PSN synchronization. Upon receiving a probe, the receiver NACKs any unreceived PSNs with the same EV. Probes are small and less likely to be dropped, and increase the chance of detecting losses before the RTO fires (Figure~\ref{fig:probe_why}).

\subsubsection{Retransmitted (RTX) Packet Probing}
Retransmitted packets' PSNs don't respect expected sequential ordering. As a result, the receiver cannot identify if retransmitted packets are lost. This motivates accompanying each RTX packet with a per-packet \textit{RTX probe}. This header-only probe carries the PSN of the retransmission. If the receiver receives the probe but not the corresponding data packet, it NACKs that PSN.

Probe loss affects latency, not correctness. If an RTX packet, probe, or NACK is lost, the sender falls back to the normal retransmission timeout. At a regular RSS update, Algorithm~\ref{alg:main} considers only subflows that returned feedback during the current interval. A completely silent subflow is therefore rerouted only when the optional no-feedback timer described in Section~\ref{subsec:reactive} is enabled. Later packets or probes on an active section can expose earlier missing PSNs. Common losses are therefore detected before timeout, while pathological combinations remain RTO-bounded.

\subsubsection{Headroom Buffer for Probe Packets}
Probe packets play an essential role in PFLD: they detect tail losses, synchronize state, and protect retransmissions. Although they are header-only packets and thus less likely to be dropped, they are enqueued with data packets and remain exposed to congestion. As an optional optimization, a deployment may reserve a small switch-buffer headroom for probes; it is not required for correctness or for the end-host-only design. Our evaluation shows that less than 0.2\% of the low-priority queue is sufficient to avoid probe drops entirely (Figure~\ref{fig:headroom_why}).

\section{Large Scale Simulations}\label{sec:sim}
We implement RSS and PFLD in the htsim simulator~\cite{htsim}, and run large-scale per-packet simulations.

\subsection{Methodology \& Setup} \label{sim_methodology}
We simulate a 2-tiered fat-tree topology with 1,024 nodes with 64-port switches.
We set the MTU to 4 KiB and the latency of all links to $1.5~\mu s$ (\textit{i.e.} a minimum RTT of $12~\mu s$ when traversing both tiers).

We compare several network settings: a symmetric case with all links having 400 Gbps capacity, and an asymmetric case with 4 out of the 16 aggregate switches having all of their links reduced to 100 Gbps.
The congestion control algorithm used is MPRDMA~\cite{mprdma}, which manages window-based congestion control on a per-flow basis. We replace only its path-selection logic with each evaluated scheme. Under RSS, MPRDMA's window is shared across all subflows; each subflow transmits at $1/\sigma$ of the aggregate rate, so RSS controls \emph{where} packets go while MPRDMA controls \emph{how fast} they are sent. Under ECMP and PLB, MPRDMA operates on a single path; under RPS and REPS, each packet is placed on a path independently.

We configure the retransmission timeout (RTO) for each experiment based on the topology's maximum unloaded RTT, with additional allowance for queueing delay. Unless noted otherwise, we set the corruption drop rate to $5 \times 10^{-5}$ (Sec.~\ref{failures}). We verified that our results hold qualitatively across rates from $10^{-6}$ to $10^{-4}$: lower rates reduce the absolute gap between schemes because fewer drops occur, while higher rates amplify PFLD's advantage because more losses require fast detection. The LLaMA experiment uses $5 \times 10^{-4}$, and the finite HPC trace replays use $5 \times 10^{-3}$ so that enough losses occur to compare recovery behavior.

\subsubsection{Simulation Workloads}\label{subsubsec:evals_workload}

We provide an extensive evaluation section with simulation-based results. We use ATLAHS~\cite{shen2025atlahsapplicationcentricnetworksimulator} with htsim to simulate advanced workloads. The workloads are summarized below:
\begin{itemize}[leftmargin=10pt]
\item \textbf{Microbenchmarks:} These are workloads classically used by previous works including Incast (many-to-one), and Tornado (one-to-one but all going across the tree to stress the \lb) traffic patterns.
\item \textbf{AI-collectives:} These include the most common collectives used for synchronous AI training: all-reduce using the ring algorithm \cite{li2020pytorch} and all-to-all \cite{naumov2019deep}.
\item \textbf{End-to-end AI traces:} We use ATLAHS~\cite{shen2025atlahsapplicationcentricnetworksimulator} to collect end-to-end traces for a single training iteration of the LLaMA 7B model~\cite{touvron2023llama}, an established benchmarking model~\cite{mlcommons_training}.
\item \textbf{HPC application traces:} We replay MPI traces from three representative HPC applications: ICON (climate modeling)~\cite{icon_pham}, LAMMPS (molecular dynamics)~\cite{thompson2022lammps}, and MILC (lattice QCD)~\cite{milc}.
\item \textbf{Data center traces:} We sample flow sizes from a Google RPC distribution at 80\% offered load to evaluate production-like message-size diversity~\cite{plb}.
\end{itemize}

\subsubsection{Schemes}
We evaluate RSS with and without PSN-based PFLD. Section~\ref{sensitivity} shows the impact on RSS of the number of \subflows and the update interval.
For all other experiments, we set the number of \subflows to $16$, and the update interval to $15~\mu s$, roughly equal to the RTT value. We compare it against the following baselines:

\begin{itemize}[leftmargin=10pt]
    \item {\textbf{ECMP~\cite{ecmp}:}} Hashes flow-level information to pick a single path for all packets in a flow. The selected path is maintained for the lifetime of the flow.
    \item {\textbf{\plb~\cite{plb}}:} Operates similarly to ECMP but proactively reroutes flows that see high ECN marking rates.
    We choose the same parameters as advised in the Flowbender~\cite{flowbender} paper (the ancestor of \plb); the ECN threshold is set to $0.05$ and flows are rerouted every congestion period.
    \item {\textbf{Flowlet Switching~\cite{flowlet}}}: Splits a flow into flowlets, bursts separated by an inter-packet gap above a threshold. Upon gap detection, the next flowlet is hashed to a different path.
    \item {\textbf{RPS~\cite{spray}}:} Randomly selects a path per packet, providing fine-grained load balancing without congestion awareness.
    \item {\textbf{REPS~\cite{bonato2025repsrecycledentropypacket}}}: A lightweight decentralized per packet adaptive load balancer. It sprays packets across multiple paths while caching high-quality paths and reusing that entropy to bias future choices. Compared to RPS, it can react to asymmetries and failures.
    \item {\textbf{RACK-TLP~\cite{rack1, tlp_draft_2013}}:} The state-of-the-art sender-side loss recovery mechanism, standardized in RFC\,8985 and used in Linux and Google's datacenter stack. We pair RACK-TLP with RSS for load balancing to approximate proprietary Falcon~\cite{falcon} architecture. This isolates the effect of PFLD's receiver-driven loss detection versus RACK-TLP's sender-side heuristics under identical load balancing.
\end{itemize}
 
\subsection{Microbenchmark Evaluation}\label{micro_eval}

We use a set of standard microbenchmarks to assess both \lb performance and loss-detection reactivity. For each workload, we evaluate the performance with and without corruption drops. Each scheme is run with and without packet trimming.
To account for randomness, we run each configuration 100 times with different random seeds and report violin plots over all flows across runs. In grey we also report a dotted line with the ideal completion time assuming perfect fairness.

In Figure~\ref{fig:incast}, we showcase an 8:1 incast where each sender transmits 1 MiB to a receiver. As expected, with packet trimming enabled and no corruption drops, schemes with finer granularity, such as REPS, RSS, and RPS, perform best. However, once trimming is disabled, only RSS+PFLD maintains the same performance because it detects losses more quickly. With corruption drops enabled, the main advantage of PFLD becomes clear. Even with trimming, corruption drops remain problematic for the other schemes, and RACK-TLP has a delayed reaction, while PFLD reacts proactively to all kinds of drops. This makes the 99th-percentile FCT with RSS+PFLD roughly half that of ECMP. PFLD's probes add a small common-case overhead, which can make its median slightly higher than RACK-TLP's, but immediate receiver-side evidence provides the larger improvement at the tail.

Then, in Figure~\ref{fig:tornado}, we focus on a Tornado pattern, where each node sends to and receives from its symmetric node in the other half of the tree. This ensures that all packets of all flows traverse the network, stressing the load balancer. The trends mirror the incast case: without corruption drops, RSS+PFLD is only slightly slower than per-packet schemes; when trimming is disabled and drops occur, RSS+PFLD gains a substantial lead, up to $2.5\times$ at the 99th.

Across both workloads, we also measured the total number of retransmission timeouts triggered. With PFLD, timeouts are reduced by 99\% compared to ECMP, both with and without trimming. In practice, the only remaining timeouts with PFLD are edge cases such as lost NACKs or simultaneous loss of a probe and all its associated data packets. This confirms that PFLD's probing mechanisms effectively eliminate RTO as the dominant recovery path, which is the root cause of the tail latency improvements observed above.

\subsection{AI Workloads}\label{sec:ai_workloads}

We show simulated results for two commonly used collectives in AI training: the all-reduce implemented via the ring algorithm~\cite{li2020pytorch}, and the all-to-all implemented using an algorithm where we limit the number of parallel connections per node~\cite{hammingmesh, naumov2019deep}. Later, we also use ATLAHS~\cite{shen2025atlahsapplicationcentricnetworksimulator} to run end-to-end training workloads using open-source traces \cite{spcl_traces_2025}.

\begin{figure}[h]
    \centering
    \includegraphics[width=0.95\linewidth]{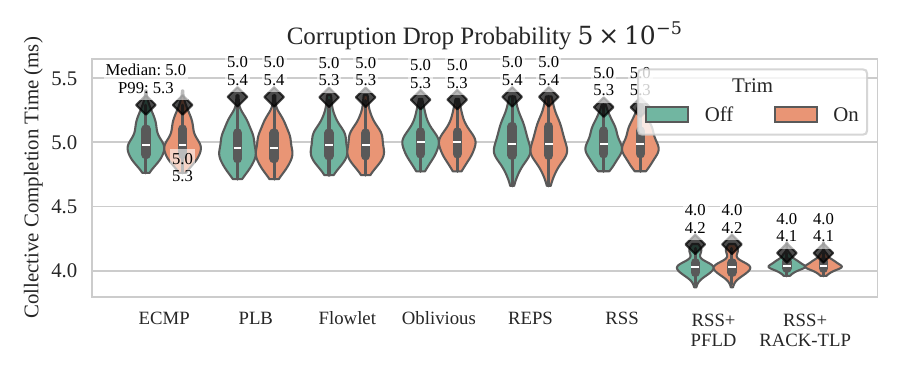}
    \caption{8 MiB ring all-reduce with $5\times10^{-5}$ corruption.}
    \label{fig:allred}
    \compactstackedfigurespace
\end{figure}

\begin{figure}[h]
    \centering
\includegraphics[width=0.95\linewidth]{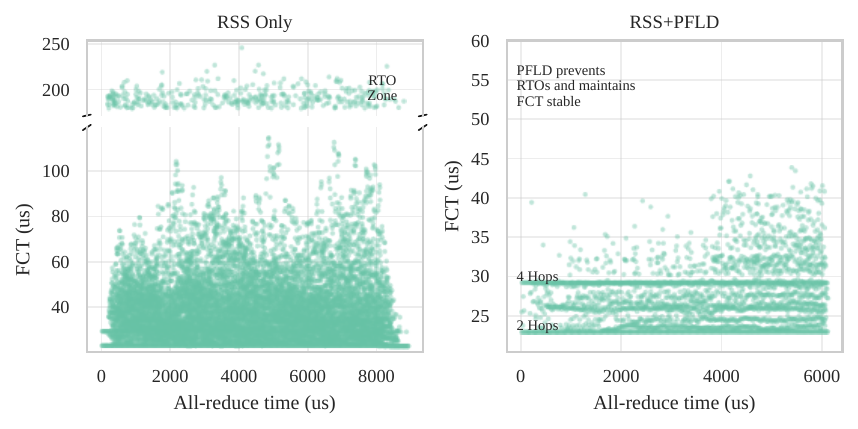}
    \caption{Individual flow completion time of all-reduce flows.}
    \label{fig:allred_fct}
    \compactfigurespace
\end{figure}

\begin{figure}[h]
    \centering
    \includegraphics[width=0.95\linewidth]{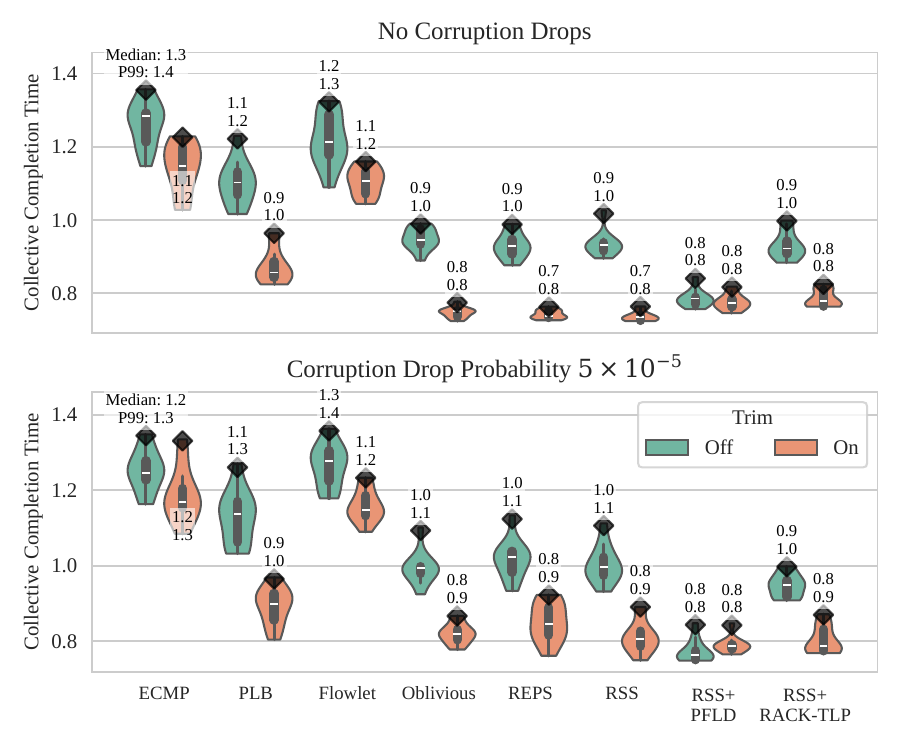}
    \caption{1 MiB all-to-all. Positive-drop panels use corruption probability $5\times10^{-5}$.}
    \label{fig:alltoall}
    \compactstackedfigurespace
\end{figure}

In Figure~\ref{fig:allred}, we show all-reduce performance with corruption drops. We omit results without corruption drops because they are indistinguishable on our fully provisioned topology. We run the all-reduce on 128 nodes 100 times and plot the distribution of collective completion times. PFLD and RACK-TLP outperform all other approaches because they provide an accurate loss-detection signal, yielding about a 35\% speedup. Their median completion times are essentially identical; at the 99th percentile, RACK-TLP is marginally faster (about 4.1\,ms versus 4.2\,ms for PFLD).

A more important benefit of PFLD is improved synchronization of the ring. Under ideal conditions, a ring all-reduce should not induce congestion. However, if a flow is slowed down, incast can arise and compound quickly. This is evident in Figure~\ref{fig:allred_fct}, which shows flow completion times over time for RSS with and without PFLD. With PFLD there is a clear pattern, with flows completing at a small set of values depending on hop count; without PFLD, the pattern is chaotic. Flows may time out, stalling part of the ring while other flows continue, which triggers incast and disrupts synchronization.

In Figure~\ref{fig:alltoall} we evaluate performance of all-to-all. Without corruption, per-packet load balancers perform slightly better than PFLD with trimming. However, without trimming, PFLD gets a 20\% lead over the second-best algorithm with and without corruption drops.

\begin{figure}[t]
    \centering
    \includegraphics[width=\columnwidth]{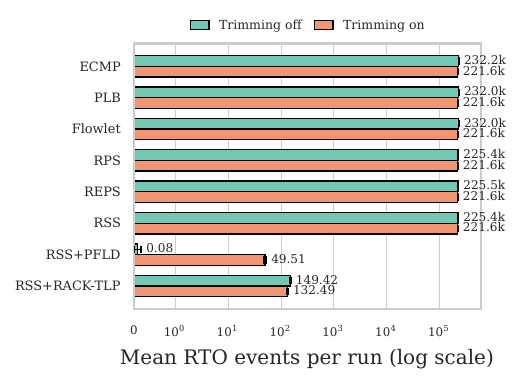}
    \caption{RTO events for a 128-node all-to-all with 4 MiB per directed flow and at most 32 active senders. Bars report means across 100 matched seeds; across-seed variation is negligible at this scale.}
    \label{fig:alltoall_rto}
\end{figure}

We also take a deeper look at RTO counts for the all-to-all experiment in Figure~\ref{fig:alltoall_rto}. RSS+PFLD averages only 0.08 RTOs per run without trimming and 49.51 with trimming, versus 149.42 and 132.49 for RSS+RACK-TLP. All other schemes exceed 221,000 RTOs per run. Thus, RSS+PFLD makes timeouts exceptional without trimming and remains well below RSS+RACK-TLP with trimming.

\begin{figure}[htbp]
    \centering
    \includegraphics[width=0.86\linewidth]{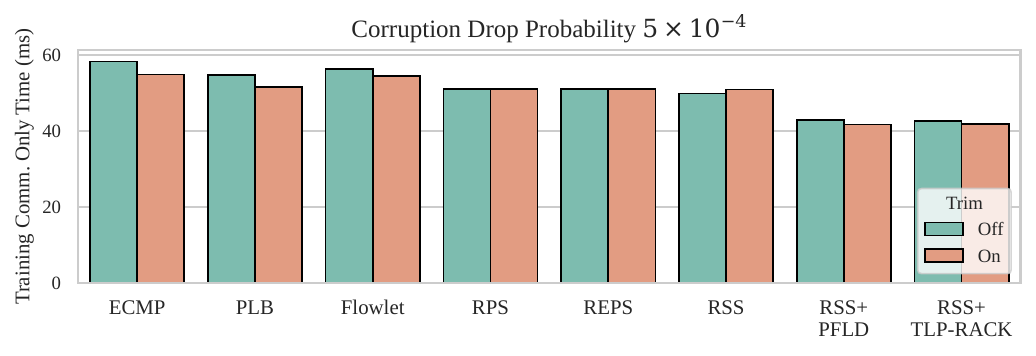}
    \caption{LLaMA 7B training: communication time per iteration at corruption probability $5\times10^{-4}$.}
    \label{fig:llama}
\end{figure}

\begin{figure*}[ht]
    \centering
    \includegraphics[width=0.95\linewidth]{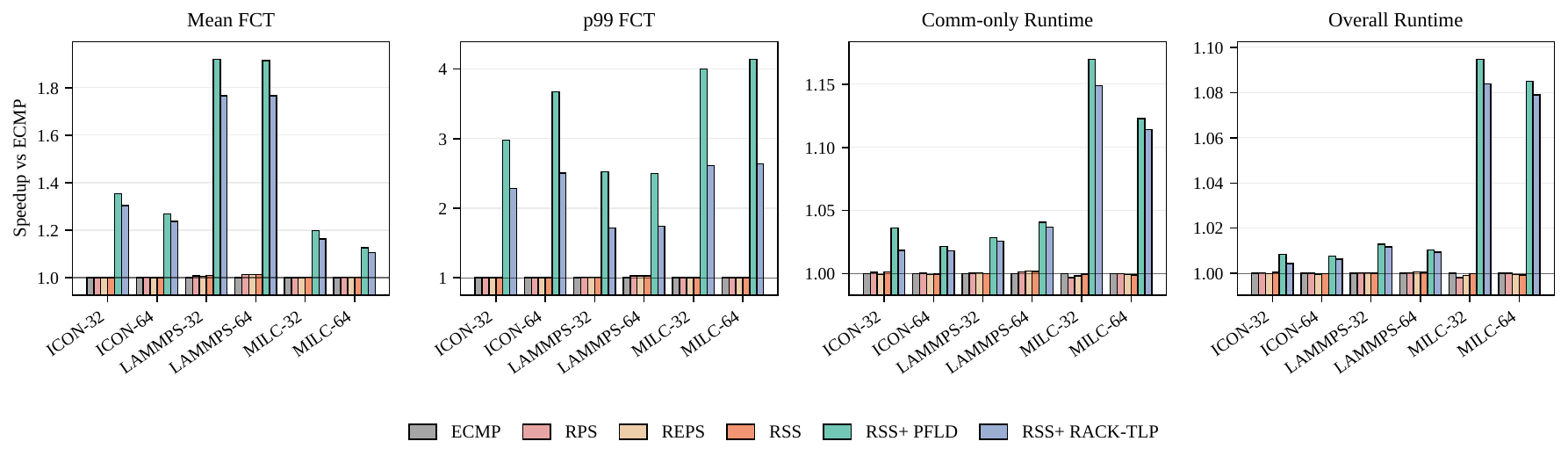}
    \caption{Performance on replayed HPC execution windows, normalized to ECMP. Real HPC traces expose limited headroom for pure load balancing, so the main gains of RSS+PFLD come from faster and more precise loss recovery, which sharply reduces FCT tails and shortens communication phases.}
    \label{fig:hpc_fig}
\end{figure*}

We also evaluate LLaMA 7B using the traces available at~\cite{spcl_traces_2025} in Figure~\ref{fig:llama}. We measure the time to complete one training iteration and plot only the exposed communication portion since the network cannot improve the compute-heavy parts. Compared to ECMP, RSS+PFLD reduces exposed communication time by 28\% and achieves a 10\% improvement over REPS and RPS. Notably, while trimming improves the performance of single-path schemes, it provides little benefit for multipath schemes.

\subsection{HPC Workloads}\label{hpc}

We evaluate RSS+PFLD on execution traces from representative HPC applications, replayed end-to-end under random corruption drops. In contrast to the synthetic tests, these traces expose smaller headroom for load-balancing gains: the standard load-balancing baselines remain relatively close in communication-only and end-to-end runtimes, while the tail FCT is significantly lower with loss recovery schemes. We report mean FCT and p99 FCT to show tail behavior, communication-only runtime to isolate the network effect, and end-to-end trace runtime to quantify the overall application impact. Across all workloads in Figure~\ref{fig:hpc_fig}, RSS+PFLD is the fastest, with RACK-TLP being second. In MILC, the most communication-heavy workload, PFLD improves tail FCT by about $1.6\times$ over RACK-TLP and communication-only runtime by up to 15\% over ECMP.

\begin{figure}[h]
  \centering
  \begin{subfigure}[b]{0.49\columnwidth}
    \centering
    \includegraphics[width=\linewidth]{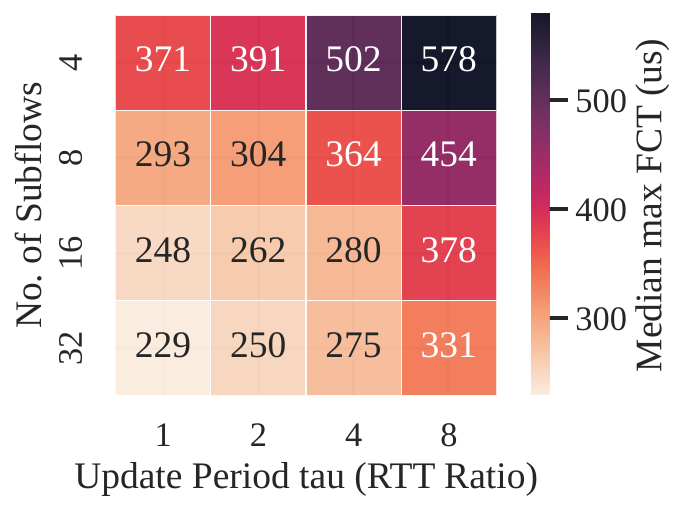}
    \caption{\symmetricCaption}
    \label{fig:evals_all2all_sym}
  \end{subfigure}\hfill
  \begin{subfigure}[b]{0.49\columnwidth}
    \centering
    \includegraphics[width=\linewidth]{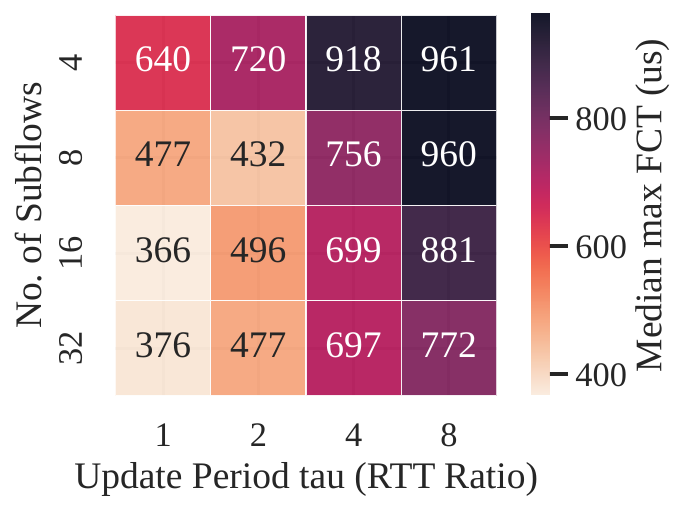}
    \caption{\asymmetricCaption}
    \label{fig:evals_all2all_asym}
  \end{subfigure}
  \caption{RSS's sensitivity to changes in $\sigma$ and $\tau$ during a tornado workload to stress its \lb capabilities.}
  \label{fig:rssGrid}
\end{figure}

\begin{figure}[!b]
    \compactfigurespace
    \centering
    \includegraphics[width=\columnwidth]{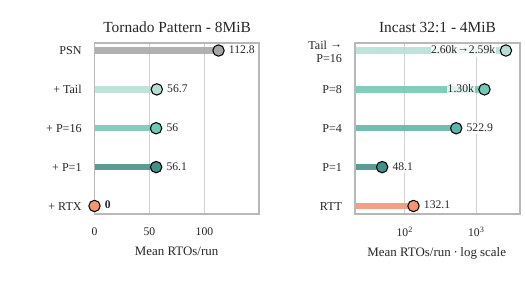}
    \caption{PFLD probe ablation. $P=k$ denotes one periodic proactive probe after every $k$ data packets; RTT denotes at most one periodic probe per RTT. Tail $\rightarrow P=16$ summarizes the nearly identical section/tail-only and $P=16$ settings, while RTX adds retransmission probes. Both workloads are run without trimming and with corruption; variation across seeds is negligible at this scale.}
    \label{fig:probe_ablation}
\end{figure}

\subsection{Data Center Traces}\label{datacenter}
\begin{figure*}[t]
    \centering
    \includegraphics[width=\textwidth]{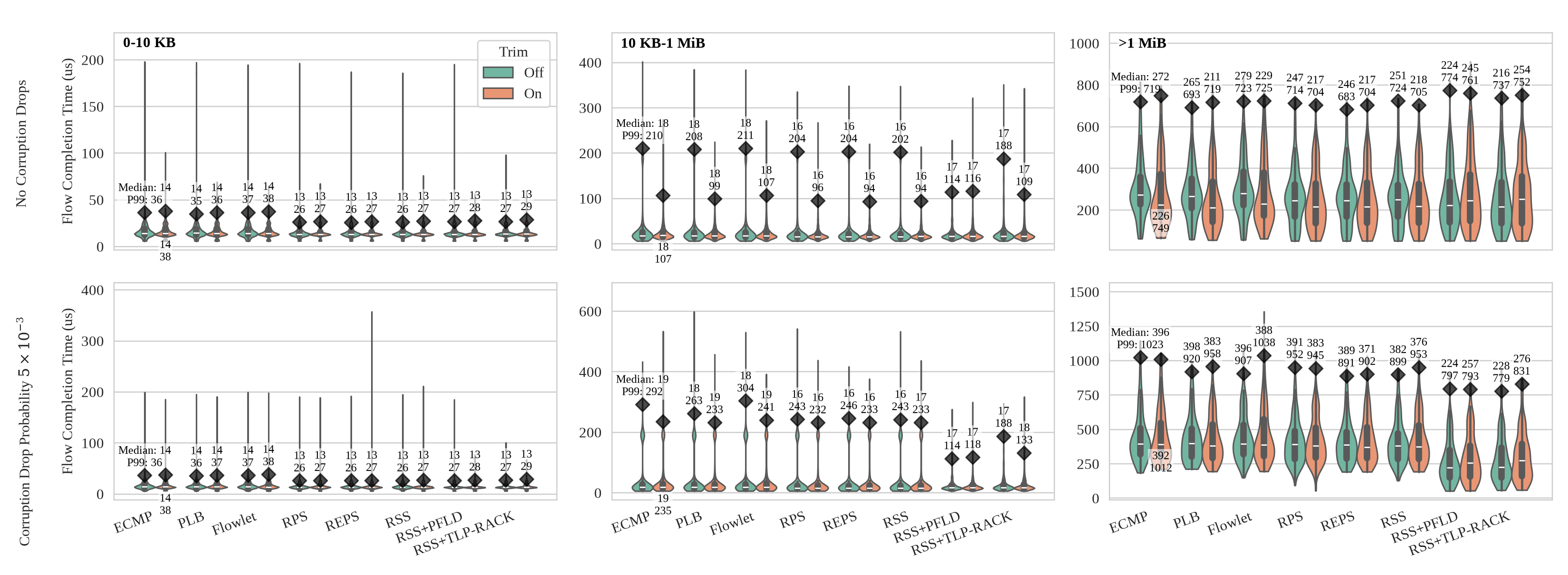}
    \caption{Flow completion times for a Google RPC workload at 80\% offered load, grouped by message size. The top row disables corruption drops and the bottom row uses corruption probability $5\times10^{-3}$. Diamonds mark the 99th percentile.}
    \label{fig:dc_trace_fct}
\end{figure*}

\begin{figure*}[t]
    \centering

    \begin{subfigure}[b]{0.48\textwidth}
        \centering
        \includegraphics[width=\linewidth]{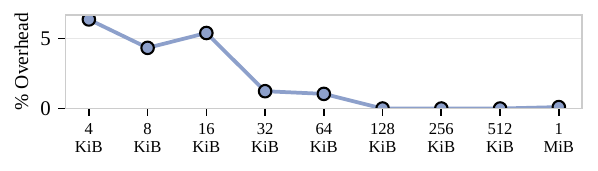}
        \caption{Performance overhead of probes.}
        \label{fig:pfld_motiv}
    \end{subfigure}\hfill
    \begin{subfigure}[b]{0.48\textwidth}
        \centering
        \includegraphics[width=\linewidth]{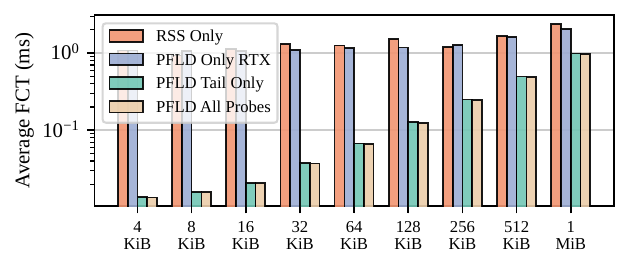}
        \caption{Observed average FCT with 1s timeout and 0.001 drop rate.}
        \label{fig:rss_motiv}
    \end{subfigure}

    \caption{PFLD DPU hardware prototype results.}
    \label{fig:hw_results}
\end{figure*}

We evaluate a workload based on Google RPC size distributions~\cite{plb}. We sample flow sizes from the empirical distribution, adjust interarrival times to reach 80\% offered load, and simulate approximately 100,000 flows. Figure~\ref{fig:dc_trace_fct} groups results into 0 to 10 KB, 10 KB to 1 MiB, and larger than 1 MiB. For the smallest messages, medians and tails remain similar because a short flow rarely contains a corruption drop. With corruption drops, RSS+PFLD roughly halves the 99th-percentile FCT of 10 KB to 1 MiB messages relative to schemes without specialized recovery. For messages larger than 1 MiB, it lowers median FCT by about 41\% relative to ECMP and remains competitive with RSS+RACK-TLP across both trimming settings.

\subsection{Parameter Sensitivity}\label{sensitivity}

In Figure~\ref{fig:rssGrid} we show the impact of RSS's update interval $\tau$ and number of \subflows $\sigma$ on the maximum FCT.
In general, adding more subflows helps improve performance, but gains are limited after a certain point (e.g., 16/32 subflows). Counterintuitively, adding subflows can actually have a negative impact in certain cases: 1) More subflows require more state. 2) More subflows in RSS make the PFLD reaction time slower as each subflow might have few packets in-flight per RTT. 3) Increasing the number of subflows means that each subflow gets less information and RSS might make uninformed routing decisions. For $\tau$, a small but reasonable (multiple of RTT) value is preferred for frequent updates.

We then run an ablation study, shown in Figure~\ref{fig:probe_ablation}, to evaluate the impact of PFLD's different probe types and frequencies. For tornado traffic, section and tail probes halve RTO events from 112.8 to 56.7; periodic probes add little in this case, while retransmission probes eliminate the remaining exposed RTOs. Under 32:1 incast, periodic frequency matters: $P=16$ remains near 2,600 RTO events, while $P=8$, $P=4$, and $P=1$ reduce the mean to 1,300, 522.9, and 48.1, respectively; probing at most once per RTT yields 132.1. Each of the 16 subflows sends infrequently during this large incast, so ordinary data packets provide sparse ordering evidence and RTT probes are especially useful. Probe type and frequency could adapt dynamically to traffic pace and network conditions. Retransmission probes instead address the residual recovery failures in the tornado workload.

\section{Hardware Feasibility Validation} \label{sec:hw}

To complement the large-scale simulation study, we validate the practical feasibility of PFLD's probing mechanisms on a programmable Data Processing Unit (DPU; specifications withheld for confidentiality). The goal of this prototype is not to replicate the full-scale evaluation of Section~\ref{sec:sim}, but to answer three research questions: (1)~does PFLD's probing logic introduce measurable overhead on real hardware? (2)~does the implementation produce spurious retransmissions? and (3)~do the different probing variants (tail, retransmission, proactive) behave on hardware as predicted by simulation?

\paragraph{Setup}
We implement RSS and PFLD on the DPU's programmable data path, maintaining receiver-side loss-detection state and probe-generation logic in the DPU's memory. The prototype runs in loopback mode, continuously sending fixed-size messages (4\,KiB to 1\,MiB) for 20 seconds per configuration. This message range represents the lower end of typical data-plane traffic and is a conservative setting for overhead evaluation, since probing costs are amortized over fewer data packets.

\paragraph{Probing overhead is negligible}
We compare a baseline RSS-only implementation against one incorporating PFLD's probing. As shown in Figure~\ref{fig:pfld_motiv}, retransmission probes add zero overhead (as expected, since no drops occur in this experiment and hence no retransmission probes are generated). Tail probes add about 6\% overhead for the smallest messages (4\,KiB, where each message is a single MTU-sized packet), and converge to zero once messages exceed 128\,KiB. This confirms that PFLD's probing is practical even on resource-constrained DPU hardware.

\paragraph{Functional correctness and probing benefits}
We emulate a corruption drop rate of 0.001 at the receiver and verify that PFLD's retransmission count exactly matches the drop count across all variants, with no spurious retransmissions. Figure~\ref{fig:rss_motiv} compares the average FCT under a 1-second RTO timeout. RSS with only out-of-sequence detection suffers frequent timeouts for tail losses, as does PFLD with only retransmission probes. Adding tail probes eliminates these timeouts, reducing average FCT by up to 2.48$\times$ for 1\,MiB messages and substantially more for smaller messages. The full PFLD configuration (all probe types) matches or improves on every variant, confirming the results from Section~\ref{pfld}.

\section{Related Work}\label{sec:relatedWork}

To maximize network utilization, the field has recently moved towards finer-grained \lb. While flow-based schemes like Flowbender~\cite{flowbender} and PLB~\cite{plb} offer improvements over ECMP, their granularity is limited. Approaches that split flows, such as flowlet switching~\cite{flowlet}, trade \lb precision for packet ordering, a compromise that is less effective for dense RDMA traffic which often lacks the idle intervals needed for rerouting~\cite{mprdma}. Per-packet schemes like RPS~\cite{spray} and DRB~\cite{drb} offer the best granularity but falter in asymmetric topologies or demand global network knowledge. DRILL~\cite{drill}, Adaptive Routing (AR) switch-based solutions~\cite{nvidia_spectrumx_platform_2025, slingshot}, SGLB~\cite{sglb}, and REPS~\cite{bonato2025repsrecycledentropypacket}, also proposed as part of UE~\cite{ultraethernet2025spec}, solve these issues to different degrees but introduce challenges: SGLB requires global switch coordination, AR demands switch modifications, and REPS struggles with ACK coalescing.

Ordered Chaos builds upon the evolution of these multipath transports by addressing key limitations in how these designs couple load balancing with loss detection. Early multipath approaches such as \emph{MPTCP} primarily target bandwidth aggregation and resilience by operating multiple underlying TCP subflows, while relying on conventional per-flow path selection (e.g., ECMP) within the network~\cite{mptcp}. Ordered Chaos extends beyond this model by introducing \rss{} subflows that share a single connection context and by explicitly structuring load balancing to preserve controlled ordering, allowing loss detection to be integrated directly into the multipath design rather than layered atop replicated transport instances.

Fine-grained \lb inherently introduces out-of-order (OOO) packet arrivals. This breaks traditional RDMA reliability mechanisms like IRN~\cite{irn}, which conflate reordering with packet loss and trigger waves of spurious retransmissions. Recognizing this, some solutions rely on switch support for explicit loss notification. Works like NDP~\cite{htsim}, CP~\cite{cp}, and more recently DCP~\cite{lossyrdma} for RDMA, use packet trimming to have the switch convert a dropped data packet into a header-only notification. Link Guardian~\cite{joshi2023linklocal} takes a complementary approach with link-local retransmission at the switch level, masking corruption losses before they reach the endpoint. While effective, these approaches require deploying new logic across the network fabric. More recently, MRC extends RoCEv2 with per-packet multipathing and selective reliability feedback over best-effort Ethernet~\cite{sohan2026mrc}; Ordered Chaos explores a complementary subflow-ordered design point. Other end-host solutions present their own trade-offs. For instance, to avoid spurious retransmissions with its Adaptive Routing, the NVIDIA SuperNIC relies on timeouts for loss recovery. Falcon~\cite{falcon} pairs subflow-based multipathing with RACK-TLP~\cite{rack1,tlp_draft_2013} for loss recovery, an architecture conceptually similar to RSS; however, RACK-TLP's sender-side heuristics degrade when subflows carry few packets. Other approaches either lack precision, like the predictive heuristics in STrack~\cite{le2024strackreliablemultipathtransport}, or trade efficiency for reordering tolerance, like RACK-TLP itself, which delays retransmission by a full round-trip time and incurs significant memory overhead. MPRDMA~\cite{mprdma} addresses multipath congestion control and is orthogonal to our work; we use it as the congestion control layer in our evaluation.

Ordered Chaos complements the RACK-TLP approach by reshaping multipath load balancing itself: \rss{} distributes traffic at subflow granularity rather than per packet, retaining most of the benefits of dynamic spraying while preserving partial flow order with in-order delivery guarantees within each subflow. This preserved structure enables \pfld{} to provide \emph{receiver-driven}, deterministic loss identification and precise recovery with minimal duplicate retransmissions, even under aggressive multipathing, and without reliance on timers, heuristic thresholds, or special switch features.

\section{Conclusion}\label{sec:conclusion}
 
The emergence of multipath-native transports, exemplified by specifications like Ultra Ethernet, signals an industry-wide shift towards best-effort out-of-order delivery to maximize network performance. However, this embrace of performance-oriented \textit{chaos} creates a critical reliability gap, as fully out-of-order transports often lack robust mechanisms for fast and precise loss detection. In this work, we presented Ordered Chaos, a unified, end-host system that directly addresses this gap. Our design combines Reactive Subflow Spraying (RSS), which provides adaptive, fine-grained load balancing using ordered subflows, with Precise Fast Loss Detection (PFLD), which leverages order within flows and adds lightweight probing to rapidly identify losses from both congestion and corruption while being end-host only deployable. Our DPU-based prototype further confirms that PFLD's probing mechanism is practical for implementation on real hardware with negligible overhead. Additionally, our evaluation across microbenchmarks, AI collectives, end-to-end LLM training, and replayed HPC application traces demonstrates that this synergy significantly improves performance: RSS+PFLD improves tail flow completion time by up to 3$\times$, reduces timeouts by up to 99\%, and speeds up collective and training communication by up to 35\%.

\section*{Acknowledgments}
This research is supported by the National Research Foundation, Singapore (NRF), and the Ministry of Digital Development and Information (MDDI) under the AI Visiting Professorship (Award No. AIVP-2025-005).  This work also received support from the ERC PSAP project (grant agreement No. 101002047). We also thank the Swiss National Supercomputing Center (CSCS) for providing the computational resources used in this work.

The authors used ChatGPT 5.4 and Claude Opus 4.6 for editing throughout the paper. All ideas and content are exclusively the original work of the authors.

\bibliographystyle{IEEEtran}
\bibliography{reference}

@article{hammingmesh,
author = {Hoefler, Torsten and Bonato, Tommaso and De Sensi, Daniele and Di Girolamo, Salvatore and Li, Shigang and Heddes, Marco and Goel, Deepak and Castro, Miguel and Scott, Steve},
title = {HammingMesh: A Network Topology for Large-Scale Deep Learning},
year = {2024},
issue_date = {December 2024},
publisher = {Association for Computing Machinery},
address = {New York, NY, USA},
volume = {67},
number = {12},
issn = {0001-0782},
url = {https://doi.org/10.1145/3623490},
doi = {10.1145/3623490},
journal = {Commun. ACM},
month = nov,
pages = {97–105},
numpages = {9}
}

@inproceedings{zhuo2017corruption,
  author    = {Danyang Zhuo and Monia Ghobadi and Ratul Mahajan and Klaus{-}Tycho F{\"o}rster and Arvind Krishnamurthy and Thomas E. Anderson},
  title     = {Understanding and Mitigating Packet Corruption in Data Center Networks},
  booktitle = {Proceedings of the ACM SIGCOMM Conference},
  year      = {2017},
  pages     = {362--375},
  address   = {Los Angeles, CA, USA},
  publisher = {ACM},
  doi       = {10.1145/3098822.3098849},
  url       = {https://doi.org/10.1145/3098822.3098849}
}

@inproceedings{guo2015pingmesh,
  author    = {Chuanxiong Guo and Lihua Yuan and Dong Xiang and Yingnong Dang and Ray Huang and Dave Maltz and Zhaoyi Liu and Vin Wang and Bin Pang and Hua Chen and Zhi{-}Wei Lin and Varugis Kurien},
  title     = {Pingmesh: A Large-Scale System for Data Center Network Latency Measurement and Analysis},
  booktitle = {Proceedings of the ACM SIGCOMM Conference},
  year      = {2015},
  pages     = {139--152},
  address   = {London, United Kingdom},
  publisher = {ACM},
  doi       = {10.1145/2785956.2787496},
  url       = {https://doi.org/10.1145/2785956.2787496}
}

@inproceedings{roy2017passive,
  author    = {Arjun Roy and Hongyi Zeng and Jasmeet Bagga and Alex C. Snoeren},
  title     = {Passive Realtime Datacenter Fault Detection and Localization},
  booktitle = {14th USENIX Symposium on Networked Systems Design and Implementation (NSDI)},
  year      = {2017},
  pages     = {595--612},
  address   = {Boston, MA, USA},
  publisher = {USENIX Association},
  url       = {https://www.usenix.org/conference/nsdi17/technical-sessions/presentation/roy}
}

@inproceedings{joshi2023linklocal,
author = {Joshi, Raj and Song, Cha Hwan and Khooi, Xin Zhe and Budhdev, Nishant and Mishra, Ayush and Chan, Mun Choon and Leong, Ben},
title = {Masking Corruption Packet Losses in Datacenter Networks with Link-local Retransmission},
year = {2023},
isbn = {9798400702365},
publisher = {Association for Computing Machinery},
address = {New York, NY, USA},
url = {https://doi.org/10.1145/3603269.3604853},
doi = {10.1145/3603269.3604853},
booktitle = {Proceedings of the ACM SIGCOMM 2023 Conference},
pages = {288–304},
numpages = {17},
location = {New York, NY, USA},
series = {ACM SIGCOMM '23}
}

@inproceedings{conweave,
author = {Song, Cha Hwan and Khooi, Xin Zhe and Joshi, Raj and Choi, Inho and Li, Jialin and Chan, Mun Choon},
title = {Network Load Balancing with In-network Reordering Support for RDMA},
year = {2023},
isbn = {9798400702365},
publisher = {Association for Computing Machinery},
address = {New York, NY, USA},
url = {https://doi.org/10.1145/3603269.3604849},
doi = {10.1145/3603269.3604849},
booktitle = {Proceedings of the ACM SIGCOMM 2023 Conference},
pages = {816–831},
numpages = {16},
location = {New York, NY, USA},
series = {ACM SIGCOMM '23}
}

@INPROCEEDINGS{spray,
  author={Dixit, Advait and Prakash, Pawan and Hu, Y. Charlie and Kompella, Ramana Rao},
  booktitle={2013 Proceedings IEEE INFOCOM}, 
  title={On the impact of packet spraying in data center networks}, 
  year={2013},
  volume={},
  number={},
  pages={2130-2138},
  doi={10.1109/INFCOM.2013.6567015},
  url={https://doi.org/10.1109/INFCOM.2013.6567015}}

@misc{ultraethernet2025spec,
  author = {{Ultra Ethernet Consortium}},
  title = {Ultra Ethernet Specification Version 1.0},
  year = {2025},
  url = {https://ultraethernet.org/wp-content/uploads/sites/20/2025/06/UE-Specification-6.11.25.pdf},
  note = {Accessed on August 29, 2025}
}

@article{bonato2025flowcutswitchinghighperformanceadaptive,
  author={Tommaso Bonato and Daniele De Sensi and Salvatore Di Girolamo and Abdulla Bataineh and David Hewson and Duncan Roweth and Torsten Hoefler},
  title={Flowcut Switching: High-Performance Adaptive Routing With In-Order Delivery Guarantees},
  journal={IEEE Transactions on Networking},
  year={2026},
  volume={34},
  pages={1974--1987},
  doi={10.1109/TON.2025.3636209},
  url={https://doi.org/10.1109/TON.2025.3636209}
}

@inproceedings{bonato2026spritz,
  author={Tommaso Bonato and Ales Kubicek and Abdul Kabbani and Ahmad Ghalayini and Maciej Besta and Torsten Hoefler},
  title={Spritz: Path-Aware Load Balancing in Low-Diameter Networks},
  booktitle={2026 IEEE International Parallel and Distributed Processing Symposium (IPDPS)},
  year={2026},
  month=may,
  pages={527--543},
  publisher={IEEE},
  doi={10.1109/IPDPS65963.2026.00052},
  url={https://doi.org/10.1109/IPDPS65963.2026.00052}
}

@misc{sohan2026mrc,
  author={Rip Sohan and Eric Spada and Eric Davis and Mark Handley and Idan Burstein and Tony Hurson and Jithin Jose and Vivek Kashyap and Rong Pan and Sayantan Sur and Sreevatsa Anantharamu and Aviv Barnea and Adrian Caulfield and Elazar Cohen and Elliot Edmunds and Yamin Friedman and Mahdieh Ghazi and Murali Guramali and Torsten Hoefler and Vipin Jain and Abdul Kabbani and Noam Katz and Yanfang Le and Charlie Mbariky and Guglielmo Morandin and Masoud Moshref and Shane O'Neil and Michael Papamichael and Jonas Pfefferle and Siva Santosh Pyla and Costin Raiciu and David Riddoch and Karen Schramm and Yuval Shpigelman and Shahaf Shuler and Shy Shyman and Raghava Sivaramu and Amin Tootoonchian and Yang Wang},
  title={The Multipath Reliable Connection ({MRC}) Transport},
  year={2026},
  month=jun,
  eprint={2606.18170},
  archivePrefix={arXiv},
  primaryClass={cs.NI},
  doi={10.48550/arXiv.2606.18170},
  url={https://arxiv.org/abs/2606.18170}
}

@misc{nvidia_bluefield3_dpu_2021_datasheet,
  author    = {{NVIDIA Corporation}},
  title     = {NVIDIA BlueField-3 DPU},
  note      = {Datasheet},
  year      = {2021},
  month     = apr,
  url       = {https://www.nvidia.com/content/dam/en-zz/Solutions/Data-Center/documents/datasheet-nvidia-bluefield-3-dpu.pdf},
  urldate   = {2025-09-16}
}

@inproceedings{zhou2020flowevent,
  author    = {Yu Zhou and Chen Sun and Hongqiang Harry Liu and Rui Miao and Shi Bai and Bo Li and Zhilong Zheng and Lingjun Zhu and Zhen Shen and Yongqing Xi and Pengcheng Zhang and Dennis Cai and Ming Zhang and Mingwei Xu},
  title     = {Flow Event Telemetry on Programmable Data Plane},
  booktitle = {Proceedings of the 2020 Annual Conference of the ACM Special Interest Group on Data Communication on the Applications, Technologies, Architectures, and Protocols for Computer Communication},
  year      = {2020},
  pages     = {76--89},
  publisher = {Association for Computing Machinery},
  doi       = {10.1145/3387514.3406214},
  url       = {https://doi.org/10.1145/3387514.3406214}
}

@inproceedings {cp,
author = {Peng Cheng and Fengyuan Ren and Ran Shu and Chuang Lin},
title = {Catch the Whole Lot in an Action: Rapid Precise Packet Loss Notification in Data Center},
booktitle = {11th USENIX Symposium on Networked Systems Design and Implementation (NSDI 14)},
year = {2014},
isbn = {978-1-931971-09-6},
address = {Seattle, WA},
pages = {17--28},
url = {https://www.usenix.org/conference/nsdi14/technical-sessions/presentation/cheng},
publisher = {USENIX Association},
month = apr
}

@inproceedings{alibaba,
author = {Qian, Kun and Xi, Yongqing and Cao, Jiamin and Gao, Jiaqi and Xu, Yichi and Guan, Yu and Fu, Binzhang and Shi, Xuemei and Zhu, Fangbo and Miao, Rui and Wang, Chao and Wang, Peng and Zhang, Pengcheng and Zeng, Xianlong and Ruan, Eddie and Yao, Zhiping and Zhai, Ennan and Cai, Dennis},
title = {Alibaba HPN: A Data Center Network for Large Language Model Training},
year = {2024},
isbn = {9798400706141},
publisher = {Association for Computing Machinery},
address = {New York, NY, USA},
url = {https://doi.org/10.1145/3651890.3672265},
doi = {10.1145/3651890.3672265},
booktitle = {Proceedings of the ACM SIGCOMM 2024 Conference},
pages = {691–706},
numpages = {16},
location = {Sydney, NSW, Australia},
series = {ACM SIGCOMM '24}
}

@inproceedings{uno,
author = {Bonato, Tommaso and Abdous, Sepehr and Kabbani, Abdul and Ghalayini, Ahmad and Gebara, Nadeen and Lam, Terry and Agarwal, Anup and Chen, Tiancheng and Yu, Zhuolong and Taranov, Konstantin and Elhaddad, Mahmoud and De Sensi, Daniele and Ghorbani, Soudeh and Hoefler, Torsten},
title = {Uno: A One-Stop Solution for Inter- and Intra-Data Center Congestion Control and Reliable Connectivity},
year = {2025},
isbn = {9798400714665},
publisher = {Association for Computing Machinery},
address = {New York, NY, USA},
url = {https://doi.org/10.1145/3712285.3759884},
doi = {10.1145/3712285.3759884},
booktitle = {Proceedings of the International Conference for High Performance Computing, Networking, Storage and Analysis},
pages = {1195–1210},
numpages = {16},
location = {
},
series = {SC '25}
}

@online{nvidia_spectrumx_platform_2025,
  author  = {{NVIDIA Corporation}},
  title   = {Spectrum-X | Ethernet Networking Platform for AI},
  year    = {2025},
  url     = {https://www.nvidia.com/en-us/networking/spectrumx/},
  urldate = {2025-09-16},
  note    = {Product page}
}

@inproceedings{lossyrdma,
author = {Li, Wenxue and Liu, Xiangzhou and Zhang, Yunxuan and Wang, Zihao and Gu, Wei and Qian, Tao and Zeng, Gaoxiong and Ren, Shoushou and Huang, Xinyang and Ren, Zhenghang and Liu, Bowen and Zhang, Junxue and Chen, Kai and Liu, Bingyang},
title = {Revisiting RDMA Reliability for Lossy Fabrics},
year = {2025},
isbn = {9798400715242},
publisher = {Association for Computing Machinery},
address = {New York, NY, USA},
url = {https://doi.org/10.1145/3718958.3750480},
doi = {10.1145/3718958.3750480},
booktitle = {Proceedings of the ACM SIGCOMM 2025 Conference},
pages = {85–98},
numpages = {14},
location = {S\~{a}o Francisco Convent, Coimbra, Portugal},
series = {SIGCOMM '25}
}

@manual{nvidia_connectx8_supernic_um_2025,
  author    = {{NVIDIA Corporation}},
  title     = {NVIDIA ConnectX-8 SuperNIC User Manual},
  year      = {2025},
  note      = {Includes official specifications},
  url       = {https://docs.nvidia.com/networking/display/nvidia-connectx-8-supernic-user-manual.pdf},
  urldate   = {2025-09-16}
}

@ARTICLE{srd,
  author={Shalev, Leah and Ayoub, Hani and Bshara, Nafea and Sabbag, Erez},
  journal={IEEE Micro}, 
  title={A Cloud-Optimized Transport Protocol for Elastic and Scalable HPC}, 
  year={2020},
  volume={40},
  number={6},
  pages={67-73},
  doi={10.1109/MM.2020.3016891},
  url={https://doi.org/10.1109/MM.2020.3016891}}

@inproceedings{falcon,
author = {Singhvi, Arjun and Dukkipati, Nandita and Chandra, Prashant and Wassel, Hassan M. G. and Sharma, Naveen Kr. and Rebello, Anthony and Schuh, Henry and Kumar, Praveen and Montazeri, Behnam and Bansod, Neelesh and Thomas, Sarin and Cho, Inho and Seibert, Hyojeong Lee and Wu, Baijun and Yang, Rui and Li, Yuliang and Huang, Kai and Yin, Qianwen and Agarwal, Abhishek and Vaduvatha, Srinivas and Wang, Weihuang and Moshref, Masoud and Ji, Tao and Wetherall, David and Vahdat, Amin},
title = {Falcon: A Reliable, Low Latency Hardware Transport},
year = {2025},
isbn = {9798400715242},
publisher = {Association for Computing Machinery},
address = {New York, NY, USA},
url = {https://doi.org/10.1145/3718958.3754353},
doi = {10.1145/3718958.3754353},
booktitle = {Proceedings of the ACM SIGCOMM 2025 Conference},
pages = {248–263},
numpages = {16},
location = {S\~{a}o Francisco Convent, Coimbra, Portugal},
series = {SIGCOMM '25}
}

@misc{hoefler2025ultraethernetsdesignprinciples,
      title={Ultra Ethernet's Design Principles and Architectural Innovations}, 
      author={Torsten Hoefler and Karen Schramm and Eric Spada and Keith Underwood and Cedell Alexander and Bob Alverson and Paul Bottorff and Adrian Caulfield and Mark Handley and Cathy Huang and Costin Raiciu and Abdul Kabbani and Eugene Opsasnick and Rong Pan and Adee Ran and Rip Sohan},
      year={2025},
      eprint={2508.08906},
      archivePrefix={arXiv},
      primaryClass={cs.NI},
      url={https://arxiv.org/abs/2508.08906}, 
}

@inproceedings {hashing_polarization,
author = {Yunhong Xu and Keqiang He and Rui Wang and Minlan Yu and Nick Duffield and Hassan Wassel and Shidong Zhang and Leon Poutievski and Junlan Zhou and Amin Vahdat},
title = {Hashing Design in Modern Networks: Challenges and Mitigation Techniques},
booktitle = {2022 USENIX Annual Technical Conference (USENIX ATC 22)},
year = {2022},
isbn = {978-1-939133-29-45},
address = {Carlsbad, CA},
pages = {805--818},
url = {https://www.usenix.org/conference/atc22/presentation/xu},
publisher = {USENIX Association},
month = jul
}

@inproceedings{sglb,
author = {Qi, Chenchen and Wu, Wenfei and Wang, Yongcan and He, Keqiang and Kao, Yu-Hsiang (Sean) and He, Zongying and Yen, Chen-Yu and Jiang, Zhuo and Luo, Feng and Anubolu, Surendra and Gao, Yanjin and Lin, Bingfeng and Ni, Wenda and Yang, Yiming and Wei, Donglin and Zhou, Boyang and Wang, Jian and Ding, Shan},
title = {SGLB: Scalable and Robust Global Load Balancing in Commodity AI Clusters},
year = {2025},
isbn = {9798400715242},
publisher = {Association for Computing Machinery},
address = {New York, NY, USA},
url = {https://doi.org/10.1145/3718958.3750527},
doi = {10.1145/3718958.3750527},
booktitle = {Proceedings of the ACM SIGCOMM 2025 Conference},
pages = {626–644},
numpages = {19},
location = {S\~{a}o Francisco Convent, Coimbra, Portugal},
series = {SIGCOMM '25}
}

@inproceedings{facebook_sigcomm,
author = {Gangidi, Adithya and Miao, Rui and Zheng, Shengbao and Bondu, Sai Jayesh and Goes, Guilherme and Morsy, Hany and Puri, Rohit and Riftadi, Mohammad and Shetty, Ashmitha Jeevaraj and Yang, Jingyi and Zhang, Shuqiang and Fernandez, Mikel Jimenez and Gandham, Shashidhar and Zeng, Hongyi},
title = {RDMA over Ethernet for Distributed Training at Meta Scale},
year = {2024},
isbn = {9798400706141},
publisher = {Association for Computing Machinery},
address = {New York, NY, USA},
url = {https://doi.org/10.1145/3651890.3672233},
doi = {10.1145/3651890.3672233},
booktitle = {Proceedings of the ACM SIGCOMM 2024 Conference},
pages = {57–70},
numpages = {14},
location = {Sydney, NSW, Australia},
series = {ACM SIGCOMM '24}
}

@INPROCEEDINGS{slingshot,
  author={De Sensi, Daniele and Di Girolamo, Salvatore and McMahon, Kim H. and Roweth, Duncan and Hoefler, Torsten},
  booktitle={SC20: International Conference for High Performance Computing, Networking, Storage and Analysis}, 
  title={An In-Depth Analysis of the Slingshot Interconnect}, 
  year={2020},
  volume={},
  number={},
  pages={1-14},
  doi={10.1109/SC41405.2020.00039},
  url={https://doi.org/10.1109/SC41405.2020.00039}}

@inproceedings{bonato2025repsrecycledentropypacket,
  author={Tommaso Bonato and Abdul Kabbani and Ahmad Ghalayini and Michael Papamichael and Mohammad Dohadwala and Lukas Gianinazzi and Mikhail Khalilov and Elias Achermann and Daniele De Sensi and Torsten Hoefler},
  title={{REPS}: Recycled Entropy Packet Spraying for Adaptive Load Balancing and Failure Mitigation},
  booktitle={Proceedings of the 21st European Conference on Computer Systems},
  series={EuroSys '26},
  year={2026},
  month=apr,
  pages={225--246},
  publisher={ACM},
  doi={10.1145/3767295.3769320},
  url={https://doi.org/10.1145/3767295.3769320}
}

@inproceedings{hedera,
author = {Al-Fares, Mohammad and Radhakrishnan, Sivasankar and Raghavan, Barath and Huang, Nelson and Vahdat, Amin},
title = {Hedera: Dynamic Flow Scheduling for Data Center Networks},
year = {2010},
publisher = {USENIX Association},
address = {San Jose, CA},
booktitle = {7th USENIX Symposium on Networked Systems Design and Implementation (NSDI 10)},
month = apr,
url={https://www.usenix.org/conference/nsdi10-0/hedera-dynamic-flow-scheduling-data-center-networks}
}

@inproceedings{htsim,
author = {Handley, Mark and Raiciu, Costin and Agache, Alexandru and Voinescu, Andrei and Moore, Andrew W. and Antichi, Gianni and W\'{o}jcik, Marcin},
title = {Re-architecting datacenter networks and stacks for low latency and high performance},
year = {2017},
isbn = {9781450346535},
publisher = {Association for Computing Machinery},
address = {New York, NY, USA},
url = {https://doi.org/10.1145/3098822.3098825},
doi = {10.1145/3098822.3098825},
booktitle = {Proceedings of the Conference of the ACM Special Interest Group on Data Communication},
pages = {29–42},
numpages = {14},
location = {Los Angeles, CA, USA},
series = {SIGCOMM '17}
}

@inproceedings{plb,
author = {Qureshi, Mubashir Adnan and Cheng, Yuchung and Yin, Qianwen and Fu, Qiaobin and Kumar, Gautam and Moshref, Masoud and Yan, Junhua and Jacobson, Van and Wetherall, David and Kabbani, Abdul},
title = {PLB: congestion signals are simple and effective for network load balancing},
year = {2022},
isbn = {9781450394208},
publisher = {Association for Computing Machinery},
address = {New York, NY, USA},
url = {https://doi.org/10.1145/3544216.3544226},
doi = {10.1145/3544216.3544226},
booktitle = {Proceedings of the ACM SIGCOMM 2022 Conference},
pages = {207–218},
numpages = {12},
location = {Amsterdam, Netherlands},
series = {SIGCOMM '22}
}

@inproceedings{mptcp,
author = {Raiciu, Costin and Barre, Sebastien and Pluntke, Christopher and Greenhalgh, Adam and Wischik, Damon and Handley, Mark},
title = {Improving datacenter performance and robustness with multipath TCP},
year = {2011},
isbn = {9781450307970},
publisher = {Association for Computing Machinery},
address = {New York, NY, USA},
url = {https://doi.org/10.1145/2018436.2018467},
doi = {10.1145/2018436.2018467},
booktitle = {Proceedings of the ACM SIGCOMM 2011 Conference},
pages = {266–277},
numpages = {12},
location = {Toronto, Ontario, Canada},
series = {SIGCOMM '11}
}

@techreport{ecmp,
author = {Hopps, C.},
title = {RFC2992: Analysis of an Equal-Cost Multi-Path Algorithm},
year = {2000},
institution = {RFC Editor},
address = {USA},
url={https://datatracker.ietf.org/doc/html/rfc2992},
doi={10.17487/RFC2992}
}

@inproceedings{flowbender,
author = {Kabbani, Abdul and Vamanan, Balajee and Hasan, Jahangir and Duchene, Fabien},
title = {FlowBender: Flow-level Adaptive Routing for Improved Latency and Throughput in Datacenter Networks},
year = {2014},
isbn = {9781450332798},
publisher = {Association for Computing Machinery},
address = {New York, NY, USA},
url = {https://doi.org/10.1145/2674005.2674985},
doi = {10.1145/2674005.2674985},
booktitle = {Proceedings of the 10th ACM International on Conference on Emerging Networking Experiments and Technologies},
pages = {149–160},
numpages = {12},
location = {Sydney, Australia},
series = {CoNEXT '14}
}

@inproceedings{juggler,
author = {Geng, Yilong and Jeyakumar, Vimalkumar and Kabbani, Abdul and Alizadeh, Mohammad},
title = {Juggler: a practical reordering resilient network stack for datacenters},
year = {2016},
isbn = {9781450342407},
publisher = {Association for Computing Machinery},
address = {New York, NY, USA},
url = {https://doi.org/10.1145/2901318.2901334},
doi = {10.1145/2901318.2901334},
booktitle = {Proceedings of the Eleventh European Conference on Computer Systems},
articleno = {20},
numpages = {16},
location = {London, United Kingdom},
series = {EuroSys '16}
}

@article{thompson2022lammps,
  title={LAMMPS-a flexible simulation tool for particle-based materials modeling at the atomic, meso, and continuum scales},
  author={Thompson, Aidan P and Aktulga, H Metin and Berger, Richard and Bolintineanu, Dan S and Brown, W Michael and Crozier, Paul S and In't Veld, Pieter J and Kohlmeyer, Axel and Moore, Stan G and Nguyen, Trung Dac and others},
  journal={Computer physics communications},
  volume={271},
  pages={108171},
  year={2022},
  publisher={Elsevier},
  doi={10.1016/j.cpc.2021.108171},
  url={https://doi.org/10.1016/j.cpc.2021.108171}
}

@article{icon_pham,
    author = {Pham, T. V. and Steger, C. and Rockel, B. and Keuler, K. and Kirchner, I. and Mertens, M. and Rieger, D. and Z\"angl, G. and Fr\"uh, B.},
    doi = {10.5194/gmd-14-985-2021},
    journal = {Geoscientific Model Development},
    number = {2},
    pages = {985--1005},
    title = {ICON in Climate Limited-area Mode (ICON release version 2.6.1): a new regional climate model},
    url = {https://doi.org/10.5194/gmd-14-985-2021},
    volume = {14},
    year = {2021}
}

@inproceedings{prestoLB,
author = {He, Keqiang and Rozner, Eric and Agarwal, Kanak and Felter, Wes and Carter, John and Akella, Aditya},
title = {Presto: Edge-based Load Balancing for Fast Datacenter Networks},
year = {2015},
isbn = {9781450335423},
publisher = {Association for Computing Machinery},
address = {New York, NY, USA},
url = {https://doi.org/10.1145/2785956.2787507},
doi = {10.1145/2785956.2787507},
booktitle = {Proceedings of the 2015 ACM Conference on Special Interest Group on Data Communication},
pages = {465–478},
numpages = {14},
location = {London, United Kingdom},
series = {SIGCOMM '15}
}

@inproceedings {flowlet,
author = {Erico Vanini and Rong Pan and Mohammad Alizadeh and Parvin Taheri and Tom Edsall},
title = {Let It Flow: Resilient Asymmetric Load Balancing with Flowlet Switching},
booktitle = {14th USENIX Symposium on Networked Systems Design and Implementation (NSDI 17)},
year = {2017},
isbn = {978-1-931971-37-9},
address = {Boston, MA},
pages = {407--420},
url = {https://www.usenix.org/conference/nsdi17/technical-sessions/presentation/vanini},
publisher = {USENIX Association},
month = mar
}

@inproceedings{drb,
author = {Cao, Jiaxin and Xia, Rui and Yang, Pengkun and Guo, Chuanxiong and Lu, Guohan and Yuan, Lihua and Zheng, Yixin and Wu, Haitao and Xiong, Yongqiang and Maltz, Dave},
title = {Per-packet load-balanced, low-latency routing for clos-based data center networks},
year = {2013},
isbn = {9781450321013},
publisher = {Association for Computing Machinery},
address = {New York, NY, USA},
url = {https://doi.org/10.1145/2535372.2535375},
doi = {10.1145/2535372.2535375},
booktitle = {Proceedings of the Ninth ACM Conference on Emerging Networking Experiments and Technologies},
pages = {49–60},
numpages = {12},
location = {Santa Barbara, California, USA},
series = {CoNEXT '13}
}

@inproceedings{drill,
author = {Ghorbani, Soudeh and Yang, Zibin and Godfrey, P. Brighten and Ganjali, Yashar and Firoozshahian, Amin},
title = {DRILL: Micro Load Balancing for Low-latency Data Center Networks},
year = {2017},
isbn = {9781450346535},
publisher = {Association for Computing Machinery},
address = {New York, NY, USA},
url = {https://doi.org/10.1145/3098822.3098839},
doi = {10.1145/3098822.3098839},
booktitle = {Proceedings of the Conference of the ACM Special Interest Group on Data Communication},
pages = {225–238},
numpages = {14},
location = {Los Angeles, CA, USA},
series = {SIGCOMM '17}
}

@inproceedings{conga,
author = {Alizadeh, Mohammad and Edsall, Tom and Dharmapurikar, Sarang and Vaidyanathan, Ramanan and Chu, Kevin and Fingerhut, Andy and Lam, Vinh The and Matus, Francis and Pan, Rong and Yadav, Navindra and Varghese, George},
title = {CONGA: distributed congestion-aware load balancing for datacenters},
year = {2014},
isbn = {9781450328364},
publisher = {Association for Computing Machinery},
address = {New York, NY, USA},
url = {https://doi.org/10.1145/2619239.2626316},
doi = {10.1145/2619239.2626316},
booktitle = {Proceedings of the 2014 ACM Conference on SIGCOMM},
pages = {503–514},
numpages = {12},
location = {Chicago, Illinois, USA},
series = {SIGCOMM '14}
}

@misc{milc,
    author =    {University of Utah},
    title =     {The MIMD Lattice Computation (MILC) Collaboration},
    howpublished = {\url{https://web.physics.utah.edu/~detar/milc/}}
}

@misc{mlcommons_training,
  author = {MLCommons},
  title = {MLCommons Training Benchmark},
  howpublished = {\url{https://mlcommons.org/benchmarks/training/}},
  year = {2024},
  note = {Accessed: 2025-09-17}
}

@article{touvron2023llama,
  title   = {LLaMA: Open and Efficient Foundation Language Models},
  author  = {Touvron, Hugo and Lavril, Thibaut and Izacard, Gautier and Martinet, Xavier and
             Lachaux, Marie-Anne and Lacroix, Timoth{\'e}e and Rozi{\`e}re, Baptiste and
             Goyal, Naman and Hambro, Eric and Azhar, Faisal and Rodriguez, Aur{\'e}lien and
             Joulin, Armand and Grave, Edouard and Lample, Guillaume},
  journal = {arXiv:2302.13971},
  year    = {2023}
}

@misc{spcl_traces_2025,
  title        = {SPCL Application Traces Repository},
  author       = {{Scalable Parallel Computing Lab (SPCL), ETH Zurich}},
  year         = {2025},
  howpublished = {\url{https://storage2.spcl.ethz.ch/traces/}},
  note         = {Directory of publicly hosted application traces (e.g., LLaMA, DLRM).},
  urldate      = {2025-09-18}
}

@inproceedings{irn,
author = {Mittal, Radhika and Shpiner, Alexander and Panda, Aurojit and Zahavi, Eitan and Krishnamurthy, Arvind and Ratnasamy, Sylvia and Shenker, Scott},
title = {Revisiting network support for RDMA},
year = {2018},
isbn = {9781450355674},
publisher = {Association for Computing Machinery},
address = {New York, NY, USA},
url = {https://doi.org/10.1145/3230543.3230557},
doi = {10.1145/3230543.3230557},
booktitle = {Proceedings of the 2018 Conference of the ACM Special Interest Group on Data Communication},
pages = {313–326},
numpages = {14},
location = {Budapest, Hungary},
series = {SIGCOMM '18}
}

@techreport{tlp_draft_2013,
  title       = {Tail Loss Probe (TLP): An Algorithm for Fast Recovery of Tail Losses},
  author      = {Nandita Dukkipati and Neal Cardwell and Yuchung Cheng and Matt Mathis},
  institution = {IETF},
  type        = {Internet-Draft},
  number      = {draft-dukkipati-tcpm-tcp-loss-probe-01},
  year        = {2013},
  month       = feb,
  note        = {Work in Progress},
  url = {https://datatracker.ietf.org/doc/draft-dukkipati-tcpm-tcp-loss-probe/}
}

@misc{rack1,
  series     = {Request for Comments},
  number     = 8985,
  howpublished = {RFC 8985},
  publisher  = {RFC Editor},
  doi        = {10.17487/RFC8985},
  url        = {https://www.rfc-editor.org/info/rfc8985},
  author     = {Yuchung Cheng and Neal Cardwell and Nandita Dukkipati and Priyaranjan Jha},
  title      = {The RACK-TLP Loss Detection Algorithm for TCP},
  year       = {2021},
  month      = feb,
  pagetotal  = {29}
}

@misc{le2024strackreliablemultipathtransport,
      title={STrack: A Reliable Multipath Transport for AI/ML Clusters}, 
      author={Yanfang Le and Rong Pan and Peter Newman and Jeremias Blendin and Abdul Kabbani and Vipin Jain and Raghava Sivaramu and Francis Matus},
      year={2024},
      eprint={2407.15266},
      archivePrefix={arXiv},
      primaryClass={cs.NI},
      url={https://arxiv.org/abs/2407.15266}, 
}

@inproceedings{mprdma,
author = {Lu, Yuanwei and Chen, Guo and Li, Bojie and Tan, Kun and Xiong, Yongqiang and Cheng, Peng and Zhang, Jiansong and Chen, Enhong and Moscibroda, Thomas},
title = {Multi-path transport for RDMA in datacenters},
year = {2018},
isbn = {9781931971430},
publisher = {USENIX Association},
address = {USA},
booktitle = {Proceedings of the 15th USENIX Conference on Networked Systems Design and Implementation},
pages = {357–371},
numpages = {15},
location = {Renton, WA, USA},
series = {NSDI'18},
url={https://www.usenix.org/conference/nsdi18/presentation/lu}
}

@article{naumov2019deep,
  title={Deep learning recommendation model for personalization and recommendation systems},
  author={Naumov, Maxim and Mudigere, Dheevatsa and Shi, Hao-Jun Michael and Huang, Jianyu and Sundaraman, Narayanan and Park, Jongsoo and Wang, Xiaodong and Gupta, Udit and Wu, Carole-Jean and Azzolini, Alisson G and others},
  journal={arXiv preprint arXiv:1906.00091},
  year={2019}
}

@article{li2020pytorch,
  title={{PyTorch} Distributed: Experiences on Accelerating Data Parallel Training},
  author={Li, Shen and Zhao, Yanli and Varma, Rohan and Salpekar, Omkar and Noordhuis, Pieter and Li, Teng and Paszke, Adam and Smith, Jeff and Vaughan, Brian and Damania, Pritam and Chintala, Soumith},
  journal={Proceedings of the VLDB Endowment},
  volume={13},
  number={12},
  pages={3005--3018},
  year={2020},
  doi={10.14778/3415478.3415530},
  url={https://doi.org/10.14778/3415478.3415530}
}

@inproceedings{shen2025atlahsapplicationcentricnetworksimulator,
      author={Siyuan Shen and Tommaso Bonato and Zhiyi Hu and Pasquale Jordan and Tiancheng Chen and Torsten Hoefler},
      title={{ATLAHS}: An Application-centric Network Simulator Toolchain for {AI}, {HPC}, and Distributed Storage},
      booktitle={Proceedings of the International Conference for High Performance Computing, Networking, Storage and Analysis},
      series={SC '25},
      year={2025},
      pages={349--367},
      publisher={Association for Computing Machinery},
      address={New York, NY, USA},
      isbn={9798400714665},
      doi={10.1145/3712285.3759838},
      url={https://doi.org/10.1145/3712285.3759838},
}

\end{document}